\documentclass[suppldata]{interact}

\usepackage{epstopdf}% To incorporate .eps illustrations using PDFLaTeX, etc.
\usepackage[caption=false]{subfig}% Support for small, `sub' figures and tables

\usepackage[numbers,sort&compress]{natbib}% Citation support using natbib.sty
\bibpunct[, ]{[}{]}{,}{n}{,}{,}% Citation support using natbib.sty
\renewcommand\bibfont{\fontsize{10}{12}\selectfont}% Bibliography support using natbib.sty
\usepackage[colorlinks=false]{hyperref}
\usepackage{multirow}
\usepackage{caption}
\usepackage{subcaption}
\theoremstyle{plain}% Theorem-like structures provided by amsthm.sty

\theoremstyle{definition}

\theoremstyle{remark}

\begin{document}

%\articletype{ARTICLE TEMPLATE}% Specify the article type or omit as appropriate

\title{A Bayesian Joint Modelling of Current Status and Current Count Data}

\author{
\name{Pavithra Hariharan\textsuperscript{a}\thanks{CONTACT Pavithra Hariharan. Email: pavithrahariharan97@gmail.com } and P. G. Sankaran\textsuperscript{a}}
\affil{\textsuperscript{a}Department of Statistics, Cochin University of Science and Technology, Cochin 682 022, Kerala, India }
}

\maketitle

\begin{abstract}
Current status censoring or case I interval censoring takes place when subjects in a study are observed just once to check if a particular event has occurred. If the event is recurring, the data are classified as current count data; if non-recurring, they are classified as current status data. Several instances of dependence of these recurring and non-recurring events are observable in epidemiology and pathology. Estimation of the degree of this dependence and identification of major risk factors 
   for the events are the major objectives of such studies. The current study proposes a Bayesian method for the joint modelling of such related events,  employing a shared frailty-based semiparametric regression model. Computational implementation makes use of an adaptive Metropolis-Hastings algorithm.   Simulation studies are put into use to show the effectiveness of the method proposed and fracture-osteoporosis data are worked through to highlight its application.	
\end{abstract}


\section{Introduction}\label{sec1}

Modelling and analysing lifetime data becomes complex when dealing with various types of censoring mechanisms. In various pathological or epidemiological studies, it is often impractical to observe the subjects multiple times due to disease transmission risks or limited time and resources. This situation gives rise to current status censoring, where each subject is examined only once to determine if a specific event has occurred. For events like death that do not repeat, where the objective is to 
study the lifetime distribution, current status censoring provides data with an observed time $U>0$ and an indicator $\delta$. Here, $\delta=1$
signifies that the event has occurred by $U$, while $\delta=0$ means it has not occurred by 
$U$. The data  $\left(U, \delta\right)$ are
referred to as current status data \cite{sun2006statistical}. Current stats data are also named as case I interval censored data, since current status censoring represents a specific form of interval censoring, with either $\left[0, U\right]$ or $\left(U, \infty\right)$ as the censoring intervals. Significant contributions to the analysis of current status data include \cite{jewell2003current,keiding1996estimation,lin1998additive} among many others. 

In the case of recurring events such as multiple episodes of infections, injuries or hospitalisations, the aim is to estimate the mean count. If the current status censoring is present, it conceals the exact times of recurrence. Consequently, the available information consists solely of the monitoring time $U$ and the number of times the recurring event has taken place by $U$, termed current count data \cite{wen2016joint}. Whenever the participants of study are monitored across several time points and counts within the panel of observation are recorded, the data become panel count data \cite{sun2013statistical,wellner2000two}. A key challenge in analysing current status and current count data lies in the fact that inferences about events must be drawn based only on information gathered from a single monitoring time.

There are several instances where the recurrence of an event is correlated with a lifetime of the study subjects. For instance, the number of fractures; a recurring event,
affects the onset of osteoporosis, a non-recurring medical condition that weakens and brittles bones, thereby heightening the risk of further fractures. This relationship between fractures and osteoporosis was studied in the 2005 fracture-osteoporosis survey held at Taiwan. Another significant example in epidemiology is the repeated occurrence of unprotected sex acts leading to Human Immunodeficiency Virus (HIV) infection. Comprehending these relationships is essential for developing preventive strategies that are critical for public health management. 

Joint modelling approaches are commonly employed to analyse two or more related outcomes. This approach is particularly beneficial when the events of interest share some common underlying process.  Although the modelling approaches for recurring events and lifetime data have been developed seperately in an extensive manner, joint modelling approaches are relatively less developed. The primary objectives of the approach are to estimate the mean count of recurrent event, determine the lifetime distribution due to non-recurrent events, identify the respective risk factors and assess the degree of dependence between these events. 
   Earlier contributions in this area include correlated marginal models presented by \cite{ghosh2003semiparametric} and a nested joint model put forward by \cite{huang2003frequency}. Later, \cite{huang2004joint} explored the relationship between the hazard of the lifetime random variable and the intensity of the recurrent event process, utilising a common subject-specific latent variable known as frailty. The literature has seen some recent developments, including those by  \cite{khan2022accelerated,rogers2016analysis,wang2023joint,xu2017joint}.
   \cite{li2019bayesian,paulon2020joint} have developed Bayesian inference methods to address this scenario. 

Under interval censoring case II, joint modelling approaches are well-established, as seen in \cite{wang2021bayesian,wang2022joint,xu2018joint}. However, methods for joint modelling with current status censoring are less explored, with leading contributions from \cite{wen2016joint,wen2018pseudo}. Moreover, Bayesian methods for jointly modelling current status and current count data are not yet developed, despite Bayesian inference methods being well-developed for current status data alone \cite{dunson2002bayesian,cui2023expectation,paulon2024bayesian}.

The objective of the current study is to develop a Bayesian methodology for joint modelling of  current count and current status data. Section \ref{sec2} proposes the model, followed by Bayesian inference procedure in Section \ref{sec3}. The efficiency and robustness of the suggested approach are investigated via simulation studies in Section \ref{sec4}. The application of the proposed method is demonstrated using a fracture-osteoporosis dataset in Section \ref{sec5}. The paper concludes providing some remarks on the current work and the scope for future works in Section \ref{sec6}. 

\section{Model}\label{sec2}
Consider a scenario where the subjects in a group are monitored for two related events, one that recurs and another that happens only once. For this scenario, let $N(t)$ be the counting process for a subject, indicating how many times the recurring event has happened within the time interval $[0,t]$, where $t>0$. Additionally, define $T>0$ as the lifetime random variable, representing the time it takes for the non-recurring event to occur for the same subject. By examining $N(t)$ and $T$ together, insights into the relationship between the frequency of the recurring event and the time taken for the non-recurring event can be obtained. 

Assume $\Lambda_{10}(t)$ represents the
non-decreasing baseline mean function of $N(t)$ and $\Lambda_{20}(t)$ signifies the baseline cumulative hazard rate of $T$.  Let 
$\omega$ be a non-negative subject-specific shared latent frailty random variable that accounts for the dependence between the two events. Suppose $\textbf{X}_{1}=(X_{11},\dots,X_{1p})'$ 
is the covariate vector affecting $N(t)$ and $\textbf{X}_{2}=(X_{21},\dots,X_{2q})'$ is the covariate vector influencing $T$. Furthermore, $\boldsymbol{\beta_{1}}=(\beta_{11},\dots,\beta_{1p})'$ and $\boldsymbol{\beta_{2}}=(\beta_{21},\dots,\beta_{2q})'$ represent the vectors of regression parameters associated with $\textbf{X}_{1}$ and $\textbf{X}_{2}$ respectively.

Consider the multiplicative shared frailty model due to \cite{wen2016joint}, wherein $N(t)$ is considered as a non-homogeneous Poisson process having the mean function 
\begin{equation}\label{2.1}
    \Lambda_{1}(t|\textbf{X}_{1}, \textbf{X}_{2})=\omega \Lambda_{10}(t) \exp(\boldsymbol{\beta}_{1}'\textbf{X}_{1})
\end{equation}
and the cumulative hazard rate of $T$ is modelled as
\begin{equation}\label{2.2}
    \Lambda_{2}(t|\textbf{X}_{1}, \textbf{X}_{2})=\omega \Lambda_{20}(t) \exp(\boldsymbol{\beta}_{2}'\textbf{X}_{2}).
\end{equation}
Here, $\omega$ has a proportional impact on $ \Lambda_{1}(t|\textbf{X}_{1}, \textbf{X}_{2})$ as well as $\Lambda_{2}(t|\textbf{X}_{1}, \textbf{X}_{2})$. This multiplicative structure suggests that the occurrence rate of both the events are inflated or deflated by the frailty variable $\omega$ and thereby accounts for the association between the events. 

Assume that $\omega$ is a gamma random variable having mean 1 and variance $\psi>0$. Employing the smoothing property of conditional expectation \cite{bhat2007modern}, the marginal mean function of $N(t)$ is derived as
\begin{equation}\label{2.3}
    \Lambda_1(t|\textbf{X}_1)=\Lambda_{10}(t) \exp(\boldsymbol{\beta}_{1}'\textbf{X}_{1})
\end{equation}
and the marginal survival function of $T$ is obtained as
\begin{equation}\label{2.4}
    S_2(t|\textbf{X}_2)=(1+\psi\Lambda_{20}(t) \exp(\boldsymbol{\beta}_{2}'\textbf{X}_{2}))^{(-1/\psi)}.
\end{equation}

A high value of frailty variance $\psi$ means that the unobserved factors have a significant impact on both the processes, leading to a higher degree of dependence between them. This modelling framework is semiparametric, as it parametrically models the covariate effects and frailty distribution, whereas the baseline mean function and the baseline cumulative hazard rate are completely unknown.

\section{Bayesian Inference Procedure}\label{sec3}
Consider the multiplicative shared frailty model given in \eqref{2.1} and \eqref{2.2}. Assume the participation of $n$ independent subjects in the study with each of them  observed just once at $U>0$. Under the current status censoring, the data from each of the subjects take the form $D=(U,\delta,N,\textbf{X}_{1},\textbf{X}_{2})$, where $N(U)=N$ and $\delta=I(T\leq U)$.
The objectives are to study the impact of $\textbf{X}_{1}$ on $N(t)$, $\textbf{X}_{2}$ on $T$, the degree of dependence between $N(t)$ and $T$, and estimation of $\Lambda_{10}(t)$ and $\Lambda_{20}(t)$ using merely the current count and current status data $D$.  

Given $(\omega,\textbf{X}_{1}, \textbf{X}_{2})$, independence of $N(t)$ and $T$ is assumed. Further it is presumed that $U$ and $(\omega,N(t),T)$ are independent when conditioned on $(\textbf{X}_{1}, \textbf{X}_{2})$ and that the conditional distribution of $U$ given $(\textbf{X}_{1}, \textbf{X}_{2})$ is free from the parameters of interest $\boldsymbol{\Theta}=(\Lambda_{10}(\cdot),\Lambda_{20}(\cdot),\boldsymbol{\beta}_{1},\boldsymbol{\beta}_{2},\psi)$.  The expected likelihood $\textup{L}(\boldsymbol{\Theta}|\mathbb{D})$ on having the independent and identically distributed data $\mathbb{D}=\left\{D_i=(U,\delta_i,N_i,\textbf{X}_{1i},\textbf{X}_{2i}):i=1,\dots,n\right\}$ is
\begin{equation}\footnotesize
\begin{aligned}\label{3.1}
%\begin{multline}
\textup{L}(\boldsymbol{\Theta}|\mathbb{D} )&=\prod_{i=1}^{n}\textup{E}_{\omega_i}\left [ P(N(U_i)=N_i) P(T_i\leq U_i)^{\delta _i}P(T_i> U_i)^{1-\delta _i}|(\omega_i,\textbf{X}_{1i}, \textbf{X}_{2i})\right ]\\
&\propto \prod_{i=1}^{n}\textup{E}_{\omega_i}\big[ e^{-\omega_i \Lambda_{10}(U_i) 
e^{\boldsymbol{\beta}_{1}'\textbf{X}_{1i}}}\left(\omega_i \Lambda_{10}(U_i)
e^{\boldsymbol{\beta}_{1}'\textbf{X}_{1i}}\right)^{N_i} \left(1-e^{-\omega_i \Lambda_{20}(U_i) e^{\boldsymbol{\beta}_{2}'\textbf{X}_{2i}}}\right)^{\delta _i}e^{-\omega_i \Lambda_{20}(U_i) e^{\boldsymbol{\beta}_{2}'\textbf{X}_{2i}}(1-\delta _i)}|(\omega_i,\textbf{X}_{1i}, \textbf{X}_{2i})\big]\\
&=\prod_{i=1}^{n} \frac{\Gamma{(N_i+\psi^{-1})}}{\Gamma{\psi^{-1}}}\left [ \psi \Lambda_{10}(U_i)e^{\boldsymbol{\beta}_{1}'\textbf{X}_{1i}} \right ]^{N_i}\left [\left(1+\psi e^{\boldsymbol{\beta}_{1}'\textbf{X}_{1i}} \Lambda_{10}(U_i)+\psi e^{\boldsymbol{\beta}_{2}'\textbf{X}_{2i}} \Lambda_{20}(U_i) \right )^{-N_i-\psi^{-1}}   \right ]^{1-\delta_i}\\
&\times \left [ \left ( 1+\psi e^{\boldsymbol{\beta}_{1}'\textbf{X}_{1i}} \Lambda_{10}(U_i) \right )^{-N_i-\psi^{-1}}-  \left ( 1+\psi e^{\boldsymbol{\beta}_{1}'\textbf{X}_{1i}} \Lambda_{10}(U_i)+\psi e^{\boldsymbol{\beta}_{2}'\textbf{X}_{2i}} \Lambda_{20}(U_i) \right )^{-N_i-\psi^{-1}} \right ]^{\delta_i}.
%\end{multline}
\end{aligned}
\end{equation}
The objective is to estimate $\Lambda_{10}(t)$, $\Lambda_{20}(t)$, $\boldsymbol{\beta}_{1}$, $\boldsymbol{\beta}_{2}$, and $\psi$ by developing a Bayesian inference procedure.

\subsection{Prior Distributions}\label{subsec3.1}
A distinctive feature of Bayesian inference is the use of priors to express probabilistic beliefs about unknown parameters before considering any data. 

Let $0=v_{0}<v_{1}<v_{2}<\dots<v_{n'}$ symbolise the distinct monitoring times among $U_{i}$;$i=1,\dots,n$. As it can be observed, the likelihood function in \eqref{3.1} is influenced by the values of $\Lambda_{10}(\cdot)$ and $\Lambda_{20}(\cdot)$ at $U_i$'s. Therefore, a piecewise constant rate function is assumed for $N(t)$, with rate $\phi_{d}>0$ in the interval $(v_{d-1},v_{d}]$ for $d=1,\dots,n'$. Consequently, $\Lambda_{10}(t)$ becomes a piecewise linear mean function, 
\begin{equation}\label{3.2}
    \Lambda_{10}(t)=\sum_{d=1}^{n'}\phi_{d}\Delta_{d}(t)=\sum_{d=1}^{n'}e^{\phi_{d}^{*}}\Delta_{d}(t),
\end{equation}
where $\phi_{d}^{*}=\log(\phi_{d})$ and $\Delta_{d}(t)=\min(v_{d},t)-\min(v_{d-1},t); d=1,\dots,n'$ \cite{cook2007statistical}. In practice, the parameters $\phi_{1},\dots,\phi_{n'}$ are essentially independent. Accordingly, the prior distribution of $\boldsymbol{\phi^{*}}=(\phi_1^{*},\dots,\phi_{n'}^{*})'$ is chosen as $n'$-variate normal with mean $\boldsymbol{\varphi}=(\varphi_{1},\dots,\varphi_{n'})'$ and a diagonal matrix of dimension $n'$ as the variance covariance matrix, so as to indicate the independence. i.e. $\boldsymbol{\phi^{*}}\sim N_{n'}(\boldsymbol{\varphi},\Sigma_{\boldsymbol{\phi^{*}}})$.

The baseline cumulative hazard rate $\Lambda_{20}(t)$, being a non-negative and non-decreasing function, is reconsidered as a step function with jumps of magnitude $\exp(\nu_{d});\nu_{d}\in \mathbb{R
}$ at $v_{d}$ for $d=1,\dots,n'$. Thereby, $\Lambda_{20}(t)$ takes the form
\begin{equation}\label{3.3}
   \Lambda_{20}(t)=-\log \left[\prod_{d:v_{d}\leq t}
    \exp\left(-e^{\nu_{d}}\right)\right],
\end{equation}
where $\nu_{d}=\log\left [ -\log\left ( \frac{S_{20}(v_{d})}{S_{20}(v_{d-1})} \right ) \right ]$ for $d=1,\dots,n'$ and $S_{20}(t)=\prod_{d:v_{d}\leq t}\exp\left(-e^{\nu_{d}}\right)$ \cite{sun2006statistical}. $N_{n'}(\boldsymbol{\vartheta},\boldsymbol{\Sigma_{\boldsymbol{\nu}}})$ is opted as the prior of 
 $\boldsymbol{\nu}=(\nu_1,\dots,\nu_{n'})'$, where $\boldsymbol{\vartheta}=(\vartheta_{1},\dots,\vartheta_{n'})'$ and  $\boldsymbol{\Sigma_{\boldsymbol{\nu}}}$,  an $n'\times n'$ matrix with non-zero
off-diagonal elements, which account for the dependence of $\nu_d$ with the   components next to it, for $d=1,\dots,n'$. %Thus, $\boldsymbol{\nu} \sim N_{n'}%%(\boldsymbol{\vartheta},\boldsymbol{\Sigma_{\boldsymbol{\nu}}})$ with the probability density function:
%\begin{equation*}
%  \pi_{\boldsymbol{\nu}}(\boldsymbol{\nu})=\frac{1}{(2\pi)^{n'/2}}|\boldsymbol{\Sigma_{\boldsymbol{\nu}}}|^{-1/2}e^{\frac{-1}{2}(\boldsymbol{\nu}-\boldsymbol{\vartheta})'\boldsymbol{\Sigma_{\boldsymbol{\nu}}}^{-1}(\boldsymbol{\nu}-\boldsymbol{\vartheta})},
% \end{equation*}

$N_{p}(\boldsymbol{\theta_{1}},\boldsymbol{\Sigma_{\boldsymbol{\beta_{1}}}})$ and $N_{q}(\boldsymbol{\theta_{2}},\boldsymbol{\Sigma_{\boldsymbol{\beta_{2}}}})$ are chosen as the priors of the real valued regression vectors $\boldsymbol{\beta_{1}}$ and $\boldsymbol{\beta_{2}}$, respectively. Here, $\boldsymbol{\theta_{1}}=(\theta_{11},\dots,\theta_{1p})'$, $\boldsymbol{\theta_{2}}=(\theta_{21},\dots,\theta_{2q})'$, and 
$\boldsymbol{\Sigma_{\boldsymbol{\beta_{1}}}}$ and $\boldsymbol{\Sigma_{\boldsymbol{\beta_{2}}}}$ are diagonal matrices entailing the independence of components within the regression vectors. Considering $\psi$ is non-negative, a normal prior is assigned for $\psi^{*}=\log(\psi)$ having mean $\chi$ and variance $\sigma^{2}$. Furthermore, $\boldsymbol{\phi^{*}}$, $\boldsymbol{\nu}$, $\boldsymbol{\beta_{1}}$, $\boldsymbol{\beta_{2}}$, and $\psi^{*}$ are assumed to be independent given the covariate vectors $\textbf{X}_{1}$ and $\textbf{X}_{2}$.

% Assuming the independence and real support of regression coefficients, a $(k+1)$-variate normal distribution is adopted as the prior for $\boldsymbol{\theta}$.  This distribution has mean vector $\boldsymbol{\tau}=(\tau_{0}, \tau_{1},\dots,\tau_{k})'$ and a $(k+1) \times (k+1)$ variance-covariance matrix $\boldsymbol{\Sigma_{\boldsymbol{\theta}}}$, having zeroes as non-diagonal elements, signifying independence between the coefficients. Therefore, $\boldsymbol{\theta} \sim N_{k+1}(\boldsymbol{\tau},\boldsymbol{\Sigma}_{\boldsymbol{\theta}})$ characterised by its probability density function
% \begin{equation*}
%  \pi_{\boldsymbol{\theta}}(\boldsymbol{\theta})=\frac{1}{(2\pi)^{(k+1)/2}}|\boldsymbol{\Sigma_{\boldsymbol{\theta}}}|^{-1/2}e^{\frac{-1}{2}(\boldsymbol{\theta}-\boldsymbol{\tau})'\boldsymbol{\Sigma_{\boldsymbol{\theta}}}^{-1}(\boldsymbol{\theta}-\boldsymbol{\tau})}, 
% \end{equation*}
% where $\boldsymbol{\tau}$ and $\boldsymbol{\Sigma_{\boldsymbol{\theta}}}$ are chosen by the experimenter to incorporate prior knowledge or uncertainty regarding the regression parameters.

%\newline
\textbf{Remark 2.1}\
The selection of hyperparameters in the prior distributions;  $\boldsymbol{\varphi}$, $\boldsymbol{\Sigma_{\boldsymbol{\phi^{*}}}}$, $\boldsymbol{\vartheta}$, $\boldsymbol{\Sigma_{\boldsymbol{\nu}}}$, $\boldsymbol{\theta_{1}}$, $\boldsymbol{\Sigma_{\boldsymbol{\beta_{1}}}}$, $\boldsymbol{\theta_{2}}$, $\boldsymbol{\Sigma_{\boldsymbol{\beta_{2}}}}$, $\chi$, and $\sigma^{2}$ depends on their informativeness. 
Informative priors use existing knowledge in Bayesian analysis, while non-informative priors stay neutral when prior knowledge is lacking. Vague priors, with large variances, allow the data to primarily shape the posterior distribution when prior information is unavailable.

\subsection{Posterior Computation}\label{subsec3.2}
Two sources of knowledge, the foundational description of the parameters laid down through priors and the likelihood obtained in the light of the data are synthesised in a natural way using Bayes’ theorem, to form the posterior distribution. 

The expected likelihood \eqref{3.1} in terms of $\boldsymbol{\Theta}=(\Lambda_{10}(\cdot), \Lambda_{20}(\cdot), \boldsymbol{\beta}_{1}, \boldsymbol{\beta}_{2}, \psi)$ is rewritten in terms of $\boldsymbol{\Theta}^{*}=(\boldsymbol{\phi^{*}}, \boldsymbol{\nu}, \boldsymbol{\beta_1}, \boldsymbol{\beta_2}, \psi^{*})$ as
{\footnotesize
\begin{multline}\label{3.4}
%\begin{split}
\textup{L}(\boldsymbol{\Theta}^{*} |\mathbb{D})=\prod_{i=1}^{n} \frac{\Gamma{(N_i+e^{-\psi^{*})}
)}}{\Gamma{e^{-\psi^{*}}}}
\left[ e^{\psi^{*}} \left(\sum_{d=1}^{n'}e^{\phi_{d}^{*}}\Delta_{d}(U_i)\right)e^{\boldsymbol{\beta}_{1}'\textbf{X}_{1i}} \right ]^{N_i}
\times
\Bigg[ 1+e^{\psi^{*}}\left(\sum_{d=1}^{n'}e^{\phi_{d}^{*}}\Delta_{d}(U_i)\right) e^{\boldsymbol{\beta_1}'\textbf{X}_{1i}}+\\
e^{\psi^{*}}\left(-\log\left[\prod_{d:v_{d}\leq U_i}
    \exp\left(-e^{\nu_{d}}\right)\right]\right) e^{\boldsymbol{\beta_2}'\textbf{X}_{2i}}\Bigg ]^{(-N_i-e^{-\psi^{*}})(1-\delta_i)}
\times \Bigg [\left ( 1+e^{\psi^{*}}\left(\sum_{d=1}^{n'}e^{\phi_{d}^{*}}\Delta_{d}(U_i)\right) e^{\boldsymbol{\beta_1}'\textbf{X}_{1i}} \right )^{-N_i-e^{-\psi^{*}}}\\
-\ \Bigg ( 1+e^{\psi^{*}}\left(\sum_{d=1}^{n'}e^{\phi_{d}^{*}}\Delta_{d}(U_i)\right) e^{\boldsymbol{\beta_1}'\textbf{X}_{1i}}+
e^{\psi^{*}}\left(-\log\left[\prod_{d:v_{d}\leq U_i}
    \exp\left(-e^{\nu_{d}}\right)\right]\Bigg) e^{\boldsymbol{\beta_2}'\textbf{X}_{2i}} \right )^{-N_i-e^{-\psi^{*}}}\Bigg ]^{\delta_i}.
\end{multline}}
Representing the prior densities of $\boldsymbol{\phi^{*}}$, $\boldsymbol{\nu}$, $\boldsymbol{\beta_{1}}$, $\boldsymbol{\beta_{2}}$, and $\psi^{*}$ respectively by $\pi(\boldsymbol{\phi^{*}})$, $\pi(\boldsymbol{\nu})$, $\pi(\boldsymbol{\beta_{1}})$, 
$\pi(\boldsymbol{\beta_{2}})$, and $\pi(\boldsymbol{\psi^{*}})$, the posterior density is constructed as
\begin{equation}\label{3.5}
\pi ^{*}(\boldsymbol{\Theta}^{*}|\mathbb{D})
\propto \textup{L}(\boldsymbol{\Theta}^{*}|\mathbb{D})\pi(\boldsymbol{\phi^{*}})\pi(\boldsymbol{\nu})\pi(\boldsymbol{\beta_{1}})\pi(\boldsymbol{\beta_{2}})\pi(\psi^{*}).
\end{equation}
The squared error loss function is widely used as the loss function for point estimation. In the Bayesian context, the Bayes estimator under squared error loss function is posterior mean of the parameter, as it minimises the Bayes risk \cite{rohatgi2015introduction}.

 Using the Bayes estimators $\Tilde{\phi}_{d}; d=1,\dots,n'$ derived by \eqref{A.1} and \eqref{A.2} in Appendix \ref{A}, an approach for estimating $\Lambda_{10}(t)$; the baseline mean function, is proposed through the formulation;
\begin{equation}\label{3.6}
    \Tilde \Lambda_{10}(t)=\sum_{d=1}^{n'}\Tilde{\phi}_{d}\Delta_{d}(t),
\end{equation}
where $\Delta_{d}(t)=\min(v_{d},t)-\min(v_{d-1},t); d=1,\dots,n'$. An estimator for the baseline cumulative hazard rate $\Lambda_{20}(t)$, utilising $\Tilde{\nu}_{d}; d=1,\dots,n'$ derived through  \eqref{A.3} and \eqref{A.4}, is suggested by
\begin{equation}\label{3.7}
    \Tilde \Lambda_{20}(t)=-\log \left[\prod_{d:v_{d}\leq t}
    \exp\left(-e^{\Tilde{\nu}_{d}}\right)\right].
\end{equation}
%\newpage
The Bayes estimators of the regression vectors $\Tilde{\boldsymbol{\beta}_{1}}=(\Tilde{\beta}_{11}, \dots, \Tilde{\beta}_{1p})'$ and $\Tilde{\boldsymbol{\beta}_{2}}=(\Tilde{\beta}_{21}, \dots, $ $\Tilde{\beta}_{2q})'$ are derived through \eqref{A.5} - \eqref{A.8}.
 $\Tilde{\psi}$ deduced by \eqref{A.9} and \eqref{A.10} is proposed as the Bayes estimator of  $\psi$, an estimator of  dependence between the recurring and non-recurring events. Based on the estimators defined so far, an estimator for marginal mean function of $N(t)$ \eqref{2.3} is proposed as
 \begin{equation}\label{3.8}
    \Tilde{\Lambda}_1(t|\textbf{X}_1)=\Tilde{\Lambda}_{10}(t) \exp(\Tilde{\boldsymbol{\beta}}_{1}'\textbf{X}_{1})
\end{equation}
and for the marginal survival function of $T$ \eqref{2.4} as 
\begin{equation}\label{3.9}
    \Tilde{S}_2(t|\textbf{X}_2)=(1+\Tilde{\psi}\Tilde{\Lambda}_{20}(t) \exp(\Tilde{\boldsymbol{\beta}}_{2}'\textbf{X}_{2}))^{(-1/\Tilde{\psi})}.
\end{equation}

\subsection{Posterior Simulation}\label{subsec3.3}
The complex forms of the proposed Bayes estimators $\Tilde{\phi}_{d}$ and $\Tilde{\nu}_{d}$ for $d=1,\dots,n'$, $\Tilde{\beta}_{1j};j=1,\dots,p$, $\Tilde{\beta}_{2j};j=1,\dots,q$, and $\Tilde{\psi}$ motivates the utilisation of Markov chain Monte Carlo (MCMC) methods for their evaluation. Since the marginal posterior densities of the parameters, elicited by \eqref{A.1}, \eqref{A.3}, \eqref{A.5}, \eqref{A.7}, and \eqref{A.10} do not possess closed forms, the scope of Gibbs sampling is ruled out. For this reason, an adaptive Metropolis algorithm (MH) due to \cite{haario1999adaptive} is employed for posterior computation. The algorithm is a powerful tool in Bayesian inference and other applications requiring efficient sampling from complex distributions. Its wide acceptance stems from its ability to well-tune the candidate distribution based on past samples, enhancing efficiency and convergence rates. Additionally, its straightforward implementation using the $\textit{MHadaptive}$ package in $R$ software has further increased its popularity. Implementation of the algorithm laid out below, facilitates the generation of Markov chain
$\boldsymbol{\Theta^{*}}^{(s)}=\big(\boldsymbol{\phi^{*}}^{(s)},\boldsymbol{\nu}^{(s)},\boldsymbol{\beta}_{1}^{(s)},\boldsymbol{\beta}_{2}^{(s)}
\psi^{*(s)} \big)$ possessing $\pi^{*}(\cdot)$ as the approximate stationary distribution, wherein $\boldsymbol{\phi^{*}}^{(s)}=(\phi_1^{* (s)},\dots,\phi_{n'}^{*(s)})$,
$\boldsymbol{\nu}^{(s)}=(\nu_{1}^{(s)},\dots,\nu_{n'}^{(s)})'$, 
$\boldsymbol{\beta_{1}}^{(s)}=(\beta_{11}^{(s)},\dots,\beta_{1p}^{(s)})'$,  and $\boldsymbol{\beta_{2}}^{(s)}=(\beta_{21}^{(s)},\dots,\beta_{2q}^{(s)})'$. \\\\
(i) Formulate function \eqref{3.5} by applying appropriate priors to the parameters of interest while considering the data $\mathbb{D}$.\\
%\textit{\underline{Step 2}}\\
(ii) Put in place the initial parameter values $\boldsymbol{\Theta}^{*(0)}$ and figure out the Maximum A posteriori (MAP) estimates $\boldsymbol{\Theta}^{*(1)}$ that
maximise \eqref{3.5}. Then, set $s=1$.\\
(iii) Pick out the Gaussian candidate distribution, using the inverse of the observed Fisher information matrix evaluated at the MAP estimates as the variance-covariance matrix and the MAP estimates as the mean.\\
(iv) Generate new parameter values $\boldsymbol{\Theta}^{*(s)c}=\big(\boldsymbol{\phi^{*}}^{(s)c},\boldsymbol{\nu}^{(s)c},\boldsymbol{\beta_{1}}^{(s)c},\boldsymbol{\beta_{2}}^{(s)c},\psi^{*(s)c}\big)$ from the candidate distribution and randomly select $u$ from a standard uniform distribution.\\
(v) Calculate the transition probability $\textup{P} (\boldsymbol{\Theta}^{*(s)},\boldsymbol{\Theta}^{*(s)c})$ as the minimum of 1 and $\frac{\pi^{*}(\boldsymbol{\Theta}^{*(s)c}|\mathbb{D})}{\pi^{*}(\boldsymbol{\Theta}^{*(s)}|\mathbb{D})}$. If $\log\,\,u$ is smaller than or equal to $\textup{P} (\boldsymbol{\Theta}^{*(s)},\boldsymbol{\Theta}^{*(s)c})$, update $\boldsymbol{\Theta}^{*(s+1)}=\boldsymbol{\Theta}^{*(s)c}$. Otherwise, let $\boldsymbol{\Theta}^{*(s+1)}=\boldsymbol{\Theta}^{*(s)}$.\\
(vi) Increment $s$ by one and execute steps (iii)-(v) for a specified number of iterations, based on Markov chain diagnostics. At certain intervals, adaptively update the variance-covariance matrix of the candidate distribution using a portion of previously generated values \cite{haario1999adaptive}.\\
(vii) After a suitable burn-in period and thinning process, the resulting parameter values $\boldsymbol{\Theta}^{*(s)}$;
$s=1,\dots,s_0$, form a nearly independent sample from a stationary distribution,
approximating  $\pi^{*}(\cdot)$.\\
(viii) Calculate the Bayes estimators as 
\begin{equation}\label{3.10}
 \Tilde{\phi}_{d}=\frac{1}{s_0}\sum_{s=1}^{s_0}e^{\phi_{d}^{*(s)}}, d=1,\dots,n',
 \end{equation}
where $\phi_{d}^{*(s)}$ represents the $d^{th}$ element of the vector $\boldsymbol{\phi}^{*(s)}$ for $s=1,\dots, s_0$. 
\begin{equation}\label{3.11}
 \Tilde{\nu}_{d}=\frac{1}{s_0}\sum_{s=1}^{s_0}\nu_{d}^{(s)}, d=1,\dots,n',
 \end{equation}
where $\nu_{d}^{(s)}$ represents the $d^{th}$ element of the vector $\boldsymbol{\nu}^{(s)}$ for $s=1,\dots, s_0$. 
\begin{equation}\label{3.12}
 \Tilde{\beta}_{1j}=\frac{1}{s_0}\sum_{s=1}^{s_0}\beta_{1j}^{(s)}, j=1,\dots,p,
 \end{equation}
where $\beta_{1j}^{(s)}$ represents the $j^{th}$ element of the vector $\boldsymbol{\beta_{1}}^{(s)}$ for $s=1,\dots, s_0$. 
\begin{equation}\label{3.13}
 \Tilde{\beta}_{2j}=\frac{1}{s_0}\sum_{s=1}^{s_0}\beta_{2j}^{(s)}, j=1,\dots,q,
 \end{equation}
where $\beta_{2j}^{(s)}$ represents the $j^{th}$ element of the vector $\boldsymbol{\beta_{2}}^{(s)}$ for $s=1,\dots, s_0$. 
\begin{equation}\label{3.14}
 \Tilde{\psi}=\frac{1}{s_0}\sum_{s=1}^{s_0}e^{\Tilde{\psi}^{*(s)}}.
 \end{equation}
The empirical averages \eqref{3.10}, \eqref{3.11}, \eqref{3.12}, \eqref{3.13}, and \eqref{3.14} converge to integrals \eqref{A.2}, \eqref{A.4}, \eqref{A.6}, \eqref{A.8}, and \eqref{A.9} respectively.

\subsection{Model Comparison and Validation}\label{subsec3.4}
When comparing alternative models on a particular dataset, there are various approaches available in Bayesian survival analysis that help to identify the model which best matches the data. Deviance Information Criterion (DIC) \cite{geisser1979predictive} and Logarithm of Pseudo-Marginal Likelihood (LPML) \cite{spiegelhalter2002bayesian} are two among the popular model choice criteria. \cite{mahanta2021application,rahmati2022compound,zhang2014bayesian} have compared frailty-based Bayesian survival models using these measures.

Define $dev(\boldsymbol{\Theta}^{*})=-2\textup{L}(\boldsymbol{\Theta^{*}|\mathbb{D}})$ as the deviance. The DIC criterion is computed approximately using the posterior sample $\boldsymbol{\Theta}^{*(s)}$;
$s=1,\dots,s_0$ obtained by step (vii) of Subsection \ref{subsec3.3}. It is calculated using the posterior mean of the deviance; $\widetilde{dev}(\boldsymbol{\Theta}^{*})=\frac{1}{s_0}\sum_{s=1}^{s_0}dev(\boldsymbol{\Theta}^{*(s)})$ and the deviance computed at the Bayes estimates of the parameters; $dev(\widetilde{\boldsymbol{\Theta}}^{*})=dev\bigg(\frac{1}{s_0}\sum_{s=1}^{s_0}\boldsymbol{\Tilde{\phi}^{*(s)}},\frac{1}{s_0}\sum_{s=1}^{s_0}\boldsymbol{\Tilde{\nu}^{(s)}},\frac{1}{s_0}\sum_{s=1}^{s_0}\boldsymbol{\Tilde{\beta_{1}}^{(s)}},\frac{1}{s_0}\sum_{s=1}^{s_0}\boldsymbol{\Tilde{\beta_{2}}^{(s)}},\frac{1}{s_0}\sum_{s=1}^{s_0}\boldsymbol{\Tilde{\psi}^{*(s)}}\bigg)$. Specifically, the model's fit to the data is summarised by $dev(\widetilde{\boldsymbol{\Theta}}^{*})$, where smaller values indicate a better fit, whereas the effective number of parameters, $p_{D}=\widetilde{dev}(\boldsymbol{\Theta}^{*})-dev(\widetilde{\boldsymbol{\Theta}}^{*})$ conveys the model's complexity. Theory recommends that the model with the smaller value of 
\begin{equation}\label{3.15}
   \textup{DIC}=2\,\widetilde{dev}(\boldsymbol{\Theta}^{*})-dev(\widetilde{\boldsymbol{\Theta}}^{*}), 
\end{equation}
should be preferred in order to balance fit and complexity. Lower DIC value suggests that the model fits the data better while being simpler or more parsimonious.

The conditional predictive distribution from deletion of an observation is cross-validated against corresponding observed response using the Conditional Predictive Ordinate (CPO) statistic, which forms the basis of the other criterion LPML. It follows that a model performing better predictively should have a higher CPO value.  Let $\mathbb{D}^{(-i)}=\mathbb{D}-\{{D_i}\}$ denote the difference of sets $\mathbb{D}$ and $\{{D_i}\}$ and $\textup{E}_{\boldsymbol{\Theta}^{*}|\mathbb{D}}$ represent the expectation with regard to the posterior density $\pi^{*}(\boldsymbol{\Theta^{*}}|\mathbb{D})$. Therefore the $i^{th}$ CPO can be expressed as the following. 
 If $\delta_i=1$,
\begin{equation}\label{3.16}
\begin{aligned}
\textup{CPO}_i &=P\big(T_i\in (0,U_i],N(U_i)=N_i|\mathbb{D}^{(-i)}\big) \\ 
 &= \bigg(\textup{E}_{\boldsymbol{\Theta}^{*}|\mathbb{D}}\bigg[\frac{1}{P\big(T_i\in (0,U_i],N(U_i)=N_i|\boldsymbol{\Theta^{*}}\big)}\bigg]\bigg)^{-1}\\
&\approx\left(\frac{1}{s_{0}}\sum_{s=1}^{s_{0}}\left[\frac{1}{P(T_{i}\in (0,U_{i}],N(U_i)=N_i|\boldsymbol{\Theta^{*(s)}})}\right]\right)^{-1},
\end{aligned}
\end{equation}
with $P\big(T_{i}\in(0,U_{i}],N(U_i)=N_i|\boldsymbol{\Theta^{*(s)}}\big)$ computed by
{\footnotesize
\begin{multline*}
   \frac{\Gamma{(N_i+e^{-\psi^{*(s)}}
)}}{\Gamma{e^{-\psi^{*(s)}}}}
\bigg[ e^{\psi^{*(s)}} \left(\sum_{d=1}^{n'}e^{\phi_{d}^{*(s)}}\Delta_{d}(U_i)\right)e^{\boldsymbol{\beta_{1}^{(s)}}'\textbf{X}_{1i}} \bigg ]^{N_i}\times \Bigg [\left ( 1+e^{\psi^{*(s)}}\left(\sum_{d=1}^{n'}e^{\phi_{d}^{*(s)}}\Delta_{d}(U_i)\right) e^{\boldsymbol{\beta_1^{(s)}}'\textbf{X}_{1i}} \right )^{-N_i-e^{-\psi^{*(s)}}}
\\
-\ \Bigg ( 1+e^{\psi^{*(s)}}\left(\sum_{d=1}^{n'}e^{\phi_{d}^{*(s)}}\Delta_{d}(U_i)\right) e^{\boldsymbol{\beta_1^{(s)}}'\textbf{X}_{1i}}+
e^{\psi^{*(s)}}\left(-\log\left[\prod_{d:v_{d}\leq U_i}
    \exp\left(-e^{\nu_{d}^{(s)}}\right)\right]\right) e^{\boldsymbol{\beta_2^{(s)}}'\textbf{X}_{2i}} \Bigg )^{-N_i-e^{-\psi^{*(s)}}}\Bigg ].
\end{multline*}}
If $\delta_i=0$,
\begin{equation}\label{3.17}
\begin{aligned}
\textup{CPO}_i &=P\big(T_i\in (U_i,\infty),N(U_i)=N_i|\mathbb{D}^{(-i)}\big) \\ 
 &= \bigg(\textup{E}_{\boldsymbol{\Theta}^{*}|\mathbb{D}}\bigg[\frac{1}{P\big(T_i\in (U_i,\infty),N(U_i)=N_i|\boldsymbol{\Theta^{*}}\big)}\bigg]\bigg)^{-1}\\
&\approx\left(\frac{1}{s_{0}}\sum_{s=1}^{s_{0}}\left[\frac{1}{P(T_{i}\in (U_{i},\infty),N(U_i)=N_i|\boldsymbol{\Theta^{*(s)}})}\right]\right)^{-1},
\end{aligned}
\end{equation}
with $P(T_{i}\in(U_{i},\infty),N(U_i)=N_i|\boldsymbol{\Theta^{*(s)}})$ computed by
{\footnotesize
\begin{multline*}
   \frac{\Gamma{(N_i+e^{-\psi^{*(s)}}
)}}{\Gamma{e^{-\psi^{*(s)}}}}
\bigg[ e^{\psi^{*(s)}} \left(\sum_{d=1}^{n'}e^{\phi_{d}^{*(s)}}\Delta_{d}(U_i)\right)e^{\boldsymbol{\beta_{1}^{(s)}}'\textbf{X}_{1i}} \bigg ]^{N_i}\times \Bigg [
\ \Bigg ( 1+e^{\psi^{*(s)}}\left(\sum_{d=1}^{n'}e^{\phi_{d}^{*(s)}}\Delta_{d}(U_i)\right) e^{\boldsymbol{\beta_1^{(s)}}'\textbf{X}_{1i}}+\\
e^{\psi^{*(s)}}\left(-\log\left[\prod_{d:v_{d}\leq U_i}
    \exp\left(-e^{\nu_{d}^{(s)}}\right)\right]\right) e^{\boldsymbol{\beta_2^{(s)}}'\textbf{X}_{2i}} \Bigg )^{-N_i-e^{-\psi^{*(s)}}}\Bigg ].
\end{multline*}}
% A Monte Carlo approximation of $\textup{CPO}_i$ is given by 
% where,
% \begin{equation}\footnotesize
% P(\cdot)=
% \begin{cases}
% \frac{\Gamma{(N_i+e^{-\psi^{*})}
% )}}{\Gamma{e^{-\psi^{*}}}}
% \bigg[ e^{\psi^{*}} \left(\sum_{d=1}^{n'}e^{\phi_{d}^{*}}\Delta_{d}(U_i)\right)e^{\boldsymbol{\beta}_{1}'\textbf{X}_{1i}} \bigg ]^{N_i}
% \times \Bigg [\left ( 1+e^{\psi^{*}}\left(\sum_{d=1}^{n'}e^{\phi_{d}^{*}}\Delta_{d}(U_i)\right) e^{\boldsymbol{\beta_1}'\textbf{X}_{1i}} \right )^{-N_i-e^{-\psi^{*}}}
% -\\\ \Bigg ( 1+e^{\psi^{*}}\left(\sum_{d=1}^{n'}e^{\phi_{d}^{*}}\Delta_{d}(U_i)\right) e^{\boldsymbol{\beta_1}'\textbf{X}_{1i}}+
% e^{\psi^{*}}\left(-\log\left[\prod_{d:v_{d}\leq U_i}
%     \exp\left(-e^{\nu_{d}}\right)\right]\Bigg) e^{\boldsymbol{\beta_2}'\textbf{X}_{2i}} \right )^{-N_i-e^{-\psi^{*}}}\Bigg ] &; \delta_i=1 \\
%   \frac{\Gamma{(N_i+e^{-\psi^{*})}
% )}}{\Gamma{e^{-\psi^{*}}}}
% \left[ e^{\psi^{*}} \left(\sum_{d=1}^{n'}e^{\phi_{d}^{*}}\Delta_{d}(U_i)\right)e^{\boldsymbol{\beta}_{1}'\textbf{X}_{1i}} \right ]^{N_i}
% \times
% \bigg[ 1+e^{\psi^{*}}\left(\sum_{d=1}^{n'}e^{\phi_{d}^{*}}\Delta_{d}(U_i)\right) e^{\boldsymbol{\beta_1}'\textbf{X}_{1i}}\\+e^{\psi^{*}}\left(-\log\left[\prod_{d:v_{d}\leq U_i}
%     \exp\left(-e^{\nu_{d}}\right)\right]\right) e^{\boldsymbol{\beta_2}'\textbf{X}_{2i}}\bigg ]^{(-N_i-e^{-\psi^{*}})}&; \delta_i=0.  
% \end{cases}
% \end{equation}
The model's overall predictive performance is summarised through LPML computed by 
\begin{equation}\label{3.18}
\textup{LPML}=\sum_{i=1}^{n}\log\,\,\textup{CPO}_i. 
\end{equation}  
A model with a higher LPML value is more likely to be accurate.

\subsection{Bayesian Influence Diagnostics}\label{subsec3.5}
Bayesian influence diagnostics are crucial for assessing the robustness and sensitivity of Bayesian models to changes in data or model assumptions. They facilitate the identification of influential data points or potential outliers and thereby enhance the credibility of Bayesian analyses.
The well-known perturbation scheme of case deletion \cite{rd1982residuals} is used here to identify which observations are highly influential.

From a Bayesian viewpoint, all relevant information regarding $\boldsymbol{\Theta^{*}}$ is captured within the posterior distribution $\pi^{*}(\boldsymbol{\Theta^{*}}|\mathbb{D})$. The influence of $i^{th}$ observation can be measured by the discrepancy among posteriors $\pi^{*}(\boldsymbol{\Theta^{*}}|\mathbb{D})$ and $\pi^{*}(\boldsymbol{\Theta^{*}}|\mathbb{D}^{(-i)})$. The $\xi$-divergence between these two posteriors, due to \cite{peng1995bayesian}, is defined by
\begin{equation*}
D_{\xi,i}=\int_{\boldsymbol{\Theta^{*}}}\xi\bigg[\frac{\pi^{*}(\boldsymbol{\Theta^{*}}|\mathbb{D}^{(-i)})}{\pi^{*}(\boldsymbol{\Theta^{*}}|\mathbb{D})}\bigg]\pi^{*}(\boldsymbol{\Theta^{*}}|\mathbb{D})d\boldsymbol{\Theta^{*}},
\end{equation*}
with $\int_{\boldsymbol{\Theta^{*}}}$ representing the multiple integral over all parameters in $\boldsymbol{\Theta}^{*}$ and $\xi(\cdot)$ is any convex function satisfying $\xi(1)=0$. Under the choice $\xi(z)=-\log z$, $D_{\xi,i}$ becomes the Kullback-Leibler (KL) divergence $D_{\textup{KL},i}$, between $\pi^{*}(\boldsymbol{\Theta^{*}}|\mathbb{D})$ and $\pi^{*}(\boldsymbol{\Theta^{*}}|\mathbb{D}^{(-i)})$ \cite{johnson1983predictive,gelfand1991bayesian}. 
It is possible to quantify the impact of removing the $i^{th}$ case from the complete data $\mathbb{D}$ upon the joint posterior distribution $\pi^{*}(\cdot)$, using $D_{\textup{KL},i}$. In particular, $D_{\textup{KL},i}$ and $\textup{CPO}_{i}$  given in \eqref{3.16} and \eqref{3.17} are related by
 \begin{equation*}
D_{\textup{KL},i}=E_{\boldsymbol{\Theta}^{*}|\mathbb{D}}\bigg(-\log\bigg[\frac{\textup{CPO}_i}{\textup{L}_i(\boldsymbol{\Theta}^{*})}\bigg]\bigg),
 \end{equation*}
with $\textup{L}_i(\boldsymbol{\Theta}^{*})$ representing the $i^{th}$ term of the expected likelihood \eqref{3.1}. The Monte Carlo estimate of $D_{\textup{KL},i}$ is computed making use of the posterior sample $\boldsymbol{\Theta}^{*(s)}$;
$s=1,\dots,s_0$ as
\begin{equation}\label{3.19}
    D_{\textup{KL},i}\approx -\log \textup{CPO}_i+\frac{1}{s_0}\sum_{s=1}^{s_0}\log \textup{L}_i(\boldsymbol{\Theta}^{*(s)}).
\end{equation}
A plot of $D_{\textup{KL},i}$ versus $i$ gives a graphical representation of the impact of deletion of $i^{th}$ observation on the posterior. A high $D_{\textup{KL},i}$ value signals that the $i^{th}$ observation has a significant influence. The threshold of influence is determined based on the method of calibration of a divergence measure detailed in \cite{peng1995bayesian}. Small values of $D_{\textup{KL},i}$ are preferred because they indicate minimal influence of individual data points on the posterior distribution, ensuring model stability and robustness, whereas an observation is considered influential if $D_{\textup{KL},i}>$0.223 \cite{de2017bayesian}.

\section{Simulation Studies}\label{sec4}
The current section evaluates efficiency of the proposed Bayesian approach for the joint modelling of current status and current count data, through simulation studies. Model defined by \eqref{2.1} and \eqref{2.2} is considered with $p=q=1$. For each of the seven parameter combinations of $(\beta_{11}, \beta_{21}, \psi)$ being considered, 500 datasets each of size $n=500$ are generated. Followed by, $\beta_{11}$, $\beta_{21}$, $\psi$, $\Lambda_{10}(t)$, and $\Lambda_{20}(t)$ are estimated in the study.

In the simulation, two Bernoulli covariates $X_{11}$ and $X_{21}$, each with probability of success 0.5 are generated independently. To generate $U_i$ under the fixed censoring scheme, ten distinct monitoring times $v=(0.1,0.2,\dots,1)$ are fixed. Each of these distinct times are allowed to repeat $n_i$ times so that $\sum_{i=1}^{10}n_i=n$, where the $n_i$'s are determined by the generation of a random vector of size 10, from a multinomial distribution with 500 trials and 10 categories each occurring with probability 0.1.
The frailties $\omega_i;i=1,\dots,500$ are simulated from a gamma distribution of mean 1 and variance $\psi$. Letting $\Lambda_{10}(U_i)=U_i^{0.9}$, the current count $N_i$ is produced from a non-homogeneous Poisson process having mean function $\omega_i\Lambda_{10}(U_i)e^{\beta_{11}X_{11i}}$ for $i=1,\dots,500$.  The current status $\delta_i$ is simulated using a Bernoulli distribution assuming success 
probability is  $1-e^{-\omega _{i}U_i^{1.3}e^{\beta _{21}X_{21i}}}$ for $i=1,\dots,500$. The entire process produces the data $(U_i,\delta_i,N_i,X_{11i},X_{21i});i=1,\dots,n$. The data simulation is performed under a random censoring scheme as well by generating monitoring times from standard uniform distribution.

For every parameter, non-informative normal priors with high variance are employed for prioritising the data and obtaining estimates that are not subjective. $N(1,10^{2})$ is used as the prior for $\beta_{11}$, $\beta_{21}$, and $\psi^{*}$. Knowing the distinct values among $U_i$ and $\Lambda_{10}(U_i)$, the true values of $\boldsymbol{\phi}^{*}$ are computed employing the relation \eqref{3.2}. Letting these values to constitute the mean vector, a 10-variate normal prior with $\Sigma_{\boldsymbol{\phi}^{*}}=100\textup{I}_{10}$ is chosen for $\boldsymbol{\phi}^{*}$, where
$\textup{I}_{10}$ reperesents the identity matrix of order 10. Due to the dependent structure of 
$\boldsymbol{\nu}$, $N_{10}(\boldsymbol{\vartheta},\boldsymbol{\Sigma}_{10}(0.2))$ is assigned as its prior. Here, $\boldsymbol{\vartheta}=(-2.99, -2.61, -2.46, -2.4, -2.3,-2.2,$ $ -2.17, -2.13, -2.09, -2.06)'$ is computed using the relation $\vartheta_d=\log\left(-\log\left(\frac{e^{-v_{d}^{1.3}}}{e^{-v_{d-1}^{1.3}}}\right) \right)$;$ d=1,\dots,10$ and 
\begin{equation*}
\boldsymbol{\Sigma} _{n'}(\rho)=\scriptsize \begin{pmatrix}
 1&  \rho &  \rho^{2}& . & . & . & \rho^{n'-1}\\ 
 \rho&  1&  \rho&  .&  .&  .&\rho^{n'-2} \\ 
 \rho^{2}&  \rho&  1&  .& . & . & \rho^{n'-3}\\ 
 .&  .& . &  .&  &  & .\\ 
 .&  .&  .&  & . &  &. \\ 
 .&  .&  .&  &  & . & .\\
\rho^{n'-1}& \rho^{n'-2} &\rho^{n'-3}  & . & . & . &1 \\
\end{pmatrix}; 0<\rho<1.  
\end{equation*}

The adaptive MH algorithm described in Subsection \ref{subsec3.3} is used to do posterior computation, considering the data generated and the priors designed. After diagnosing the Markov chains $\boldsymbol{\Theta}^{*(s)}$ across 100,000 iterations, 10,000 samples are discarded for burn-in and the leftover samples are thinned by choosing every $30^{th}$ observation to obtain the target posterior sample. With the sample, Bayes estimates $\Tilde{\beta}_{11}$,  $\Tilde{\beta}_{21}$, and $\Tilde{\psi}$ are computed using \eqref{3.12}, \eqref{3.13}, and \eqref{3.14} respectively. One may look into Appendix \ref{B1} for further details on MCMC convergence diagnostics. An MCMC iteration takes 0.0032 s to yield a posterior observation. 
 
\setlength{\tabcolsep}{2pt} % Default value: 6pt
\renewcommand{\arraystretch}{1.0} % Default value:
% \usepackage{multirow}
\begin{table}[h!]
\centering
\caption{
The results for $(\beta_{11},\beta_{21},\psi)$ from simulation studies under scenario (1) and (2)}
\label{frequent_fixed_scenario1}
\begin{tabular}{|c|c|ccccc|ccccc|}
\hline
\multirow{2}{*}{} & \multirow{2}{*}{True} & \multicolumn{5}{c|}{(1)\;$(v_{1},\dots, v_{10})=(0.1,0.2,\dots,1.0)$}                                                                                               & \multicolumn{5}{c|}{(2)\;$v_i\sim Uniform(0,1)$}                                                                                                \\ \cline{3-12} 
                           &                             & \multicolumn{1}{c|}{Mean}    & \multicolumn{1}{c|}{Abs. bias} & \multicolumn{1}{c|}{ESD}    & \multicolumn{1}{c|}{SSE}    & CP   & \multicolumn{1}{c|}{Mean}    & \multicolumn{1}{c|}{Abs. bias} & \multicolumn{1}{c|}{ESD}    & \multicolumn{1}{c|}{SSE}    & CP   \\ \hline
$\beta_{11}$                & 0.6                        & \multicolumn{1}{c|}{0.5833} & \multicolumn{1}{c|}{0.0167}        & \multicolumn{1}{c|}{0.1324} & \multicolumn{1}{c|}{0.1443} & 0.94 & \multicolumn{1}{c|}{0.5949} & \multicolumn{1}{c|}{0.0051}        & \multicolumn{1}{c|}{0.1398} & \multicolumn{1}{c|}{0.1345} & 0.95 \\
$\beta_{21}$                & 0.8                        & \multicolumn{1}{c|}{0.7963}  & \multicolumn{1}{c|}{0.0037}        & \multicolumn{1}{c|}{0.1812} & \multicolumn{1}{c|}{ 0.1731} & 0.98 & \multicolumn{1}{c|}{0.7369}  & \multicolumn{1}{c|}{0.0631}        & \multicolumn{1}{c|}{0.1877} & \multicolumn{1}{c|}{0.2026} & 0.92 \\
$\psi$                 & 1                         & \multicolumn{1}{c|}{1.0246}  & \multicolumn{1}{c|}{0.0246}        & \multicolumn{1}{c|}{0.1473} & \multicolumn{1}{c|}{0.1576} & 0.96 & \multicolumn{1}{c|}{ 1.0450 }  & \multicolumn{1}{c|}{0.0450}        & \multicolumn{1}{c|}{0.1588} & \multicolumn{1}{c|}{0.1655} & 0.93 \\
\hline
$\beta_{11}$                & -0.9                         & \multicolumn{1}{c|}{-0.9378}        & \multicolumn{1}{c|}{0.0378   }              & \multicolumn{1}{c|}{0.1708
}       & \multicolumn{1}{c|}{0.1776}       &  0.96    & \multicolumn{1}{c|}{-0.9286}        & \multicolumn{1}{c|}{0.0286}              & \multicolumn{1}{c|}{0.1808}       & \multicolumn{1}{c|}{0.1803}       &  0.97    \\
$\beta_{21}$                & 0.8                       & \multicolumn{1}{c|}{0.7939}        & \multicolumn{1}{c|}{0.0061}              & \multicolumn{1}{c|}{0.1784}       & \multicolumn{1}{c|}{0.1788}       &  0.94    & \multicolumn{1}{c|}{0.7290}        & \multicolumn{1}{c|}{0.0709}              & \multicolumn{1}{c|}{0.1824 }       & \multicolumn{1}{c|}{0.2027}       &   0.92   \\
$\psi$                 & 0.7                      & \multicolumn{1}{c|}{0.7307}        & \multicolumn{1}{c|}{0.0307}              & \multicolumn{1}{c|}{0.1645}       & \multicolumn{1}{c|}{0.1757}       &0.94      & \multicolumn{1}{c|}{0.7400}        & \multicolumn{1}{c|}{0.0400}              & \multicolumn{1}{c|}{0.1763}       & \multicolumn{1}{c|}{0.1746}       &  0.96    \\
\hline
$\beta_{11}$                & 0.75                          & \multicolumn{1}{c|}{0.7487}        & \multicolumn{1}{c|}{0.0012}              & \multicolumn{1}{c|}{0.1219}       & \multicolumn{1}{c|}{0.1253}       &  0.93    & \multicolumn{1}{c|}{0.7452}        & \multicolumn{1}{c|}{0.0048}              & \multicolumn{1}{c|}{0.1271}       & \multicolumn{1}{c|}{0.1193}       &  0.98    \\
$\beta_{21}$                & -1.2                       & \multicolumn{1}{c|}{-1.2429}        & \multicolumn{1}{c|}{0.0429}              & \multicolumn{1}{c|}{0.2203}       & \multicolumn{1}{c|}{0.2512}       &  0.94   & \multicolumn{1}{c|}{-1.2634}        & \multicolumn{1}{c|}{0.0634}              & \multicolumn{1}{c|}{0.2331}       & \multicolumn{1}{c|}{0.2514}       &  0.93    \\
$\psi$                 & 0.6                         & \multicolumn{1}{c|}{0.6082}        & \multicolumn{1}{c|}{0.0082}              & \multicolumn{1}{c|}{0.1111}       & \multicolumn{1}{c|}{0.1186}       &    0.94  & \multicolumn{1}{c|}{0.6161}        & \multicolumn{1}{c|}{0.0161}              & \multicolumn{1}{c|}{0.1198}       & \multicolumn{1}{c|}{0.1197}       &  0.96    \\
\hline
$\beta_{11}$                & 1.1                        & \multicolumn{1}{c|}{1.1102}        & \multicolumn{1}{c|}{0.0102}              & \multicolumn{1}{c|}{0.1383}       & \multicolumn{1}{c|}{0.1434}       & 0.94     & \multicolumn{1}{c|}{1.1054}        & \multicolumn{1}{c|}{0.0054}              & \multicolumn{1}{c|}{0.1449}       & \multicolumn{1}{c|}{0.1494}       & 0.94     \\
$\beta_{21}$                & 1.3                        & \multicolumn{1}{c|}{1.2956}        & \multicolumn{1}{c|}{0.0044}              & \multicolumn{1}{c|}{0.1977}       & \multicolumn{1}{c|}{0.2047}       &0.93      & \multicolumn{1}{c|}{1.2429}        & \multicolumn{1}{c|}{0.0571}              & \multicolumn{1}{c|}{ 0.2021}       & \multicolumn{1}{c|}{0.2369}       &    0.92  \\
$\psi$                 & 1.5                        & \multicolumn{1}{c|}{1.5210}        & \multicolumn{1}{c|}{ 0.0210}              & \multicolumn{1}{c|}{0.1752}       & \multicolumn{1}{c|}{0.1647}       &  0.97   & \multicolumn{1}{c|}{1.5326}        & \multicolumn{1}{c|}{0.0326}              & \multicolumn{1}{c|}{0.1863}       & \multicolumn{1}{c|}{0.2026}       &  0.94    \\
\hline  
$\beta_{11}$                & -1                       & \multicolumn{1}{c|}{-1.0223 } & \multicolumn{1}{c|}{0.0223}        & \multicolumn{1}{c|}{ 0.1849} & \multicolumn{1}{c|}{0.1747} & 0.97 & \multicolumn{1}{c|}{-1.0123} & \multicolumn{1}{c|}{0.0123}        & \multicolumn{1}{c|}{0.1956} & \multicolumn{1}{c|}{0.1880} & 0.97\\
$\beta_{21}$                & -1.1                       & \multicolumn{1}{c|}{-1.1475}  & \multicolumn{1}{c|}{0.0475}        & \multicolumn{1}{c|}{0.2333} & \multicolumn{1}{c|}{0.2396} & 0.93& \multicolumn{1}{c|}{-1.1923}  & \multicolumn{1}{c|}{0.0923}        & \multicolumn{1}{c|}{0.2458} & \multicolumn{1}{c|}{0.2776} & 0.94 \\
$\psi$                 & 1.2                         & \multicolumn{1}{c|}{1.2364}  & \multicolumn{1}{c|}{0.0364}        & \multicolumn{1}{c|}{0.2482} & \multicolumn{1}{c|}{0.2508} & 0.94 & \multicolumn{1}{c|}{1.2669}  & \multicolumn{1}{c|}{0.0669}        & \multicolumn{1}{c|}{0.2708} & \multicolumn{1}{c|}{0.3157} & 0.92 \\
\hline
$\beta_{11}$                & -0.5                         & \multicolumn{1}{c|}{-0.5196}        & \multicolumn{1}{c|}{0.0196}              & \multicolumn{1}{c|}{0.1595}       & \multicolumn{1}{c|}{0.1697}       &  0.96    & \multicolumn{1}{c|}{-0.5408}        & \multicolumn{1}{c|}{0.0408}              & \multicolumn{1}{c|}{ 0.1678}       & \multicolumn{1}{c|}{0.1818}       &  0.93    \\
$\beta_{21}$                & 0.75                       & \multicolumn{1}{c|}{0.7374}        & \multicolumn{1}{c|}{0.0126}              & \multicolumn{1}{c|}{0.1872}       & \multicolumn{1}{c|}{ 0.1935}       &  0.94    & \multicolumn{1}{c|}{0.7162}        & \multicolumn{1}{c|}{0.0338}              & \multicolumn{1}{c|}{0.1941}       & \multicolumn{1}{c|}{0.1821}       &   0.95  \\
$\psi$                 & 1                        & \multicolumn{1}{c|}{1.0402}        & \multicolumn{1}{c|}{0.0402}              & \multicolumn{1}{c|}{0.1869}       & \multicolumn{1}{c|}{0.2069}       &0.92      & \multicolumn{1}{c|}{1.0700}        & \multicolumn{1}{c|}{0.0700}              & \multicolumn{1}{c|}{0.2036}       & \multicolumn{1}{c|}{0.2291}       &  0.94    \\
\hline
$\beta_{11}$                & 1.8                          & \multicolumn{1}{c|}{1.8045}        & \multicolumn{1}{c|}{0.0045}              & \multicolumn{1}{c|}{ 0.1234}       & \multicolumn{1}{c|}{0.1233}       &  0.95    & \multicolumn{1}{c|}{1.7923}        & \multicolumn{1}{c|}{0.0077}              & \multicolumn{1}{c|}{0.1288}       & \multicolumn{1}{c|}{0.1360}       &  0.92    \\
$\beta_{21}$                & -2                       & \multicolumn{1}{c|}{-2.1268}        & \multicolumn{1}{c|}{0.1268}              & \multicolumn{1}{c|}{0.2932}       & \multicolumn{1}{c|}{0.3355}       &  0.92   & \multicolumn{1}{c|}{-2.1446}        & \multicolumn{1}{c|}{0.1446}              & \multicolumn{1}{c|}{ 0.3192}       & \multicolumn{1}{c|}{0.3574}       &  0.94    \\
$\psi$                 & 0.95                         & \multicolumn{1}{c|}{0.9628}        & \multicolumn{1}{c|}{0.0128}              & \multicolumn{1}{c|}{0.1127}       & \multicolumn{1}{c|}{0.1249}       &    0.95 & \multicolumn{1}{c|}{0.9677}        & \multicolumn{1}{c|}{0.0177}              & \multicolumn{1}{c|}{0.1184}       & \multicolumn{1}{c|}{0.1189}       &  0.94    \\
\hline
\end{tabular}
\end{table}

\setlength{\tabcolsep}{2pt} % Default value: 6pt
\renewcommand{\arraystretch}{0.9} % Default value:
% \usepackage{multirow}
\begin{table}[h!]
\centering
\caption{
The MeanMSEs of $\Tilde{\Lambda}_{10}(t)$ under scenario (1) and (2)}
%\label{frequent_fixed_scenario2}
\label{maxmse1}
%\resizebox{14.5cm}{1.3cm}{%
\small{
\begin{tabular}{|c|c|c|c|c|c|c|c|}
\hline
                                               & (0.6,0.8,1) & (-0.9,0.8,0.7) & (0.75,-1.2,0.6) & (1.1,1.3,1.5) & (-1,-1.1,1.2) & (-0.5,0.75,1) & (1.8,-2,0.95) \\ \hline
$(1)$ & 0.0441         & 0.0607        & 0.0461         & 0.0476          & 0.0671        & 0.0550          & 0.0387          \\ 
$(2)$                         & 0.0081         & 0.0086         & 0.0069          & 0.0098         & 0.0109        &   0.0108       & 0.0402        \\ \hline
\end{tabular}}
\end{table}

\setlength{\tabcolsep}{2pt} % Default value: 6pt
\renewcommand{\arraystretch}{0.9}
\begin{table}[h!]
%\centering
\caption{
The MeanMSEs of $\Tilde{\Lambda}_{20}(t)$ under scenario (1) and (2)}
%\label{frequent_fixed_scenario2}
\label{maxmse2}
%\resizebox{14.5cm}{1.3cm}{%
\small{
\begin{tabular}{|c|c|c|c|c|c|c|c|}
\hline
                                               & (0.6,0.8,1) & (-0.9,0.8,0.7) & (0.75,-1.2,0.6) & (1.1,1.3,1.5) & (-1,-1.1,1.2) & (-0.5,0.75,1) & (1.8,-2,0.95) \\ \hline
$(1)$ & 0.0099         & 0.0098         & 0.0119          & 0.0105          & 0.0114        & 0.0105          & 0.0111          \\ 
$(2)$                         & 0.0103         & 0.0077         & 0.0107         & 0.0114       & 0.0125        & 0.0097          & 0.0119         \\ \hline
\end{tabular}}
\end{table}

Table \ref{frequent_fixed_scenario1} illustrates the frequentist operating characteristics of the estimators $\Tilde{\beta}_{11}$, $\Tilde{\beta}_{21}$, and  $\Tilde{\psi}$ for two scenarios: $(1)\,(v_{1}, \dots, v_{10})=(0.1,\dots,1)$ and $(2)\,v_{i}\sim U(0,1)$. The table includes the average of 500 posterior mean estimates (Mean), the absolute value of bias (Abs. bias), ESD computed as the mean of the standard deviations of posterior samples (PSD) and the SSE derived from the standard deviation of posterior mean estimates. Coverage probability (CP) is assessed using the 95\% Bayesian credible intervals (BCIs), indicating the proportion of BCIs that contain the true parameter value. The results indicate that under both fixed and random censoring schemes, the Mean estimates are close to the true values of $\beta_{11}$, $\beta_{21}$, and  $\psi$. Additionally, the differences between ESD and SSE are minimal and their values are low, with CP values near 0.95. These conclusions are consistent across different sample sizes, including $n=100$ and $n=250$.

$\Tilde{\Lambda}_{10}(t)$ and $\Tilde{\Lambda}_{20}(t)$ are estimated by \eqref{3.6} and \eqref{3.7}
respectively, using Bayes estimates $\Tilde{\phi}_{d}, \Tilde{\nu}_{d}; d=1,\dots,10$ computed by \eqref{3.10} and \eqref{3.11}. By calculating the local mean square errors of $\Tilde{\Lambda}_{10}(t)$ and $\Tilde{\Lambda}_{20}(t)$ at ten different monitoring times and reporting their mean values as MeanMSE in Tables \ref{maxmse1} and \ref{maxmse2}, the efficacy of the suggested method is assessed. These errors are consistently small across all settings, showing that $\Lambda_{10}(t)$ and $\Lambda_{20}(t)$ are accurately estimated.

Apart from the simulation study discussed above, an additional study has also been carried out to examine the sensitivity of the suggested method to gamma frailty assumption. A model being less sensitive to the gamma frailty assumption means that the model's estimates remain relatively stable and accurate even if the true underlying frailty distribution differs from the assumed gamma distribution. This new study relies that the frailty random variable comes from a mixture log-normal distribution $0.5 \,\textup{LN}(-0.32,0.64)+ 0.5 \,\textup{LN}(-0.125,0.25)$, having mean 1 and variance 0.79. Here, $\textup{LN}(\mu,\sigma^{2})$ refers to a log normal distribution with location parameter $\mu$ and scale parameter $\sigma^{2}$. Under this scenario, estimation procedures of $\Tilde{\beta}_{11}$, $\Tilde{\beta}_{21}$, $\Tilde{\Lambda}_{10}(t)$, and $\Tilde{\Lambda}_{20}(t)$ are carried out as proposed and the posterior summaries are tabulated in Tables \ref{misfrailty_summ}, \ref{mis_maxmse1}, and \ref{mis_maxmse2}. These efficient results further affirm that the proposed Bayesian estimation procedure is robust against violation from gamma frailty assumption.

\setlength{\tabcolsep}{2pt} % Default value: 6pt
\renewcommand{\arraystretch}{0.9} % Default value:
% \usepackage{multirow}
\begin{table}[h!]
\centering
\caption{
The results for $(\beta_{11},\beta_{21})$ under scenario (1) and (2) with misspecified frailty}
\label{misfrailty_summ}
\begin{tabular}{|c|c|ccccc|ccccc|}
\hline
\multirow{2}{*}{} & \multirow{2}{*}{True} & \multicolumn{5}{c|}{(1)\;$(v_{1}, v_{2},..., v_{10})=(0.1,0.2,\dots,1.0)$}                                                                                               & \multicolumn{5}{c|}{(2)\;$v_i\sim Uniform(0,1)$}                                                                                                \\ \cline{3-12} 
                           &                             & \multicolumn{1}{c|}{Mean}    & \multicolumn{1}{c|}{Abs. bias} & \multicolumn{1}{c|}{ESD}    & \multicolumn{1}{c|}{SSE}    & CP   & \multicolumn{1}{c|}{Mean}    & \multicolumn{1}{c|}{Abs. bias} & \multicolumn{1}{c|}{ESD}    & \multicolumn{1}{c|}{SSE}    & CP   \\ \hline
$\beta_{11}$                & 0.6                        & \multicolumn{1}{c|}{0.5870} & \multicolumn{1}{c|}{0.0129}        & \multicolumn{1}{c|}{0.1199} & \multicolumn{1}{c|}{0.1246} & 0.94 & \multicolumn{1}{c|}{0.5926} & \multicolumn{1}{c|}{0.0074}        & \multicolumn{1}{c|}{0.1261} & \multicolumn{1}{c|}{0.1318} & 0.96 \\
$\beta_{21}$                & 0.8                        & \multicolumn{1}{c|}{0.8062}  & \multicolumn{1}{c|}{0.0062}        & \multicolumn{1}{c|}{0.1631} & \multicolumn{1}{c|}{ 0.1582} & 0.95 & \multicolumn{1}{c|}{0.7532}  & \multicolumn{1}{c|}{0.0468}        & \multicolumn{1}{c|}{0.1681} & \multicolumn{1}{c|}{0.1623} & 0.96 \\
\hline
$\beta_{11}$                & -0.9                         & \multicolumn{1}{c|}{-0.9056}        & \multicolumn{1}{c|}{0.0056  }              & \multicolumn{1}{c|}{0.1630
}       & \multicolumn{1}{c|}{0.1832}       &  0.92    & \multicolumn{1}{c|}{-0.9277}        & \multicolumn{1}{c|}{0.0277}              & \multicolumn{1}{c|}{0.1708}       & \multicolumn{1}{c|}{0.2060}       &  0.92    \\
$\beta_{21}$                & 0.8                       & \multicolumn{1}{c|}{0.8059}        & \multicolumn{1}{c|}{0.0059}              & \multicolumn{1}{c|}{0.1658}       & \multicolumn{1}{c|}{0.1827}       &  0.94    & \multicolumn{1}{c|}{0.7647}        & \multicolumn{1}{c|}{0.0353}              & \multicolumn{1}{c|}{0.1686 }       & \multicolumn{1}{c|}{0.1920}       &   0.93   \\
\hline
$\beta_{11}$                & 0.75                          & \multicolumn{1}{c|}{0.7479}        & \multicolumn{1}{c|}{0.0020}              & \multicolumn{1}{c|}{0.1189}       & \multicolumn{1}{c|}{0.1245}       &  0.93    & \multicolumn{1}{c|}{0.7403}        & \multicolumn{1}{c|}{0.0097}              & \multicolumn{1}{c|}{0.1254}       & \multicolumn{1}{c|}{0.1302}       &  0.96    \\
$\beta_{21}$                & -1.2                       & \multicolumn{1}{c|}{-1.2773}        & \multicolumn{1}{c|}{0.0773}              & \multicolumn{1}{c|}{0.2175}       & \multicolumn{1}{c|}{0.2397}       &  0.92   & \multicolumn{1}{c|}{-1.3048}        & \multicolumn{1}{c|}{0.1048}              & \multicolumn{1}{c|}{0.2319}       & \multicolumn{1}{c|}{0.2635}       &  0.93    \\
\hline
$\beta_{11}$                & 1.1                        & \multicolumn{1}{c|}{1.1102}        & \multicolumn{1}{c|}{0.0102}              & \multicolumn{1}{c|}{0.1142}       & \multicolumn{1}{c|}{0.1274}       & 0.94     & \multicolumn{1}{c|}{1.0855}        & \multicolumn{1}{c|}{0.0145}              & \multicolumn{1}{c|}{0.1193}       & \multicolumn{1}{c|}{0.1095}       & 0.95     \\
$\beta_{21}$                & 1.3                        & \multicolumn{1}{c|}{1.3357}        & \multicolumn{1}{c|}{0.0357}              & \multicolumn{1}{c|}{0.1688}       & \multicolumn{1}{c|}{0.1795}       &0.95      & \multicolumn{1}{c|}{1.2706}        & \multicolumn{1}{c|}{0.0294}              & \multicolumn{1}{c|}{ 0.1734}       & \multicolumn{1}{c|}{0.1766}       &    0.95  \\
\hline  
$\beta_{11}$                & -1                       & \multicolumn{1}{c|}{-1.0189 } & \multicolumn{1}{c|}{0.0189}        & \multicolumn{1}{c|}{ 0.1682} & \multicolumn{1}{c|}{0.2016} & 0.92 & \multicolumn{1}{c|}{-1.0411} & \multicolumn{1}{c|}{0.0411}        & \multicolumn{1}{c|}{0.1695} & \multicolumn{1}{c|}{0.2036} & 0.92\\
$\beta_{21}$                & -1.1                       & \multicolumn{1}{c|}{-1.1296}  & \multicolumn{1}{c|}{0.0296}        & \multicolumn{1}{c|}{0.2075} & \multicolumn{1}{c|}{0.2172} & 0.95& \multicolumn{1}{c|}{-1.1368}  & \multicolumn{1}{c|}{0.0368}        & \multicolumn{1}{c|}{0.2123} & \multicolumn{1}{c|}{0.2691} & 0.92 \\
\hline
$\beta_{11}$                & -0.5                         & \multicolumn{1}{c|}{-0.5196}        & \multicolumn{1}{c|}{0.0196}              & \multicolumn{1}{c|}{0.1595}       & \multicolumn{1}{c|}{0.1697}       &  0.96    & \multicolumn{1}{c|}{-0.5200}        & \multicolumn{1}{c|}{0.0200}              & \multicolumn{1}{c|}{ 0.1538}       & \multicolumn{1}{c|}{0.1688}       &  0.92    \\
$\beta_{21}$                & 0.75                       & \multicolumn{1}{c|}{0.7374}        & \multicolumn{1}{c|}{0.0126}              & \multicolumn{1}{c|}{0.1872}       & \multicolumn{1}{c|}{ 0.1935}       &  0.94    & \multicolumn{1}{c|}{0.7099}        & \multicolumn{1}{c|}{0.0401}              & \multicolumn{1}{c|}{0.1701}       & \multicolumn{1}{c|}{0.1831}       &   0.94  \\
\hline
$\beta_{11}$                & 1.8                          & \multicolumn{1}{c|}{1.7846}        & \multicolumn{1}{c|}{0.0154}              & \multicolumn{1}{c|}{ 0.1075}       & \multicolumn{1}{c|}{0.1089}       &  0.96    & \multicolumn{1}{c|}{1.8108}        & \multicolumn{1}{c|}{0.0108}              & \multicolumn{1}{c|}{0.1134}       & \multicolumn{1}{c|}{0.1210}       &  0.92    \\
$\beta_{21}$                & -2                       & \multicolumn{1}{c|}{-2.0966}        & \multicolumn{1}{c|}{0.0966}              & \multicolumn{1}{c|}{0.2862}       & \multicolumn{1}{c|}{0.2853}       &  0.96   & \multicolumn{1}{c|}{-2.1931}        & \multicolumn{1}{c|}{0.1931}              & \multicolumn{1}{c|}{ 0.3177}       & \multicolumn{1}{c|}{0.3046}       &  0.92    \\
\hline

\end{tabular}
\end{table}

\setlength{\tabcolsep}{2pt} % Default value: 6pt
\renewcommand{\arraystretch}{0.9} % Default value:
% \usepackage{multirow}
\begin{table}[h!]
\centering
\caption{
The MeanMSEs of $\Tilde{\Lambda}_{10}(t)$ under scenario (1) and (2) with misspecified frailty}
%\label{frequent_fixed_scenario2}
\label{mis_maxmse1}
%\resizebox{14.5cm}{1.3cm}{%
\small{
\begin{tabular}{|c|c|c|c|c|c|c|c|}
\hline
                                               & (0.6,0.8) & (-0.9,0.8) & (0.75,-1.2) & (1.1,1.3) & (-1,-1.1) & (-0.5,0.75) & (1.8,-2) \\ \hline
$(1)$ & 0.0366         & 0.0545         & 0.0389         & 0.0321          & 0.0985        & 0.0898          & 0.1249          \\ 
$(2)$                         & 0.0072        & 0.2115         & 0.2008         & 0.0077         & 0.0203       &   0.0139       & 0.0516         \\ \hline
\end{tabular}}
\end{table}

\setlength{\tabcolsep}{2pt} % Default value: 6pt
\renewcommand{\arraystretch}{0.9}
\begin{table}[h!]
\centering
\caption{
The MeanMSEs of $\Tilde{\Lambda}_{20}(t)$ under scenario (1) and (2) with misspecified frailty}
%\label{frequent_fixed_scenario2}
\label{mis_maxmse2}
%\resizebox{14.5cm}{1.3cm}{%
\small{
\begin{tabular}{|c|c|c|c|c|c|c|c|}
\hline
                                               & (0.6,0.8) & (-0.9,0.8) & (0.75,-1.2) & (1.1,1.3) & (-1,-1.1) & (-0.5,0.75) & (1.8,-2) \\ \hline
$(1)$ & 0.0095         & 0.0099        & 0.0113          & 0.0098          & 0.0119       & 0.0093          & 0.0112         \\ 
$(2)$                         & 0.0084        & 0.0099         & 0.0108          & 0.0091        & 0.1446       &0.0094          & 0.0118         \\ \hline
\end{tabular}}
\end{table}

% The estimated curves are plotted alongside the true curves in \ref{Fig.1}.  All these demonstrate the consistent performance of the proposed joint estimation procedure. 

% \begin{figure}[h!]
% \centering
% \includegraphics[width=14cm,height=7.15cm]{Rplot03.png}
% \caption{The estimates of $F(t)$ alongside the true curve}
% \label{Fig_comparison}
% \end{figure}
%\vspace{0.5pt}

\section{An Illustration With Fracture-Osteoporosis Survey Data}\label{sec5}

Inspired by the 2005 fracture-osteoporosis survey study in Taiwan, \cite{wen2016joint,wen2018pseudo} initially undertook a statistical analysis of the data. The survey involved 1336 males and 1361 females over the age of 65, aiming to investigate factors linked to fractures and osteoporosis in the elderly. Data on the subjects' fracture and osteoporosis histories were gathered through questionnaires only once. This approach resulted in case I interval-censored data for fracture incidence (current count data) and age at onset of osteoporosis (current status data). The study's objective was to identify the risk factors for fractures and osteoporosis, considering the correlation between the two conditions. Borrowing prior information from \cite{wen2016joint} that the only significant factors are gender and tea consumption, the current section attempts to illustrate the proposed Bayesian joint estimation procedure with the data.

 The dataset considered for the current analysis consists of the variables, the age at the time of monitoring ($U$), current status of osteoporosis ($\delta$), current count of fracture ($N$), indicator of female gender 
($X_1$), and tea intake ($X_{2}$) coded from 0 to 4, with larger value denoting higher intake. The models given by \eqref{2.1} and \eqref{2.2}, with $X_1$ and $X_2$ included as covariates in both, are assumed to fit the dataset in order to obtain some prior information about the functions $\Lambda_{10}(t)$ and $\Lambda_{20}(t)$. As a result, the I-spline based maximum likelihood estimates $\hat{\Lambda}_{10}(t)$ are obtained using \textit{spef} package, while the smoothed maximum likelihood estimates  $\hat{\Lambda}_{20}(t)$ are obtained using \textit{curstatCI} package in \textit{R} software. On identifying the distinct time points at which the functions undergo a major shift, the dataset is suitably modified with thirteen distinct monitoring times $\boldsymbol{v}=(0.01, 0.02, 0.03, 0.04, 0.05, 0.09, 0.25, 0.30, 0.34, 0.38, 0.42,$ $ 0.65, 0.9)$ for the analysis, on appropriately transforming the time scale to $(0.01,1)$.

To ensure a more valuable 
analysis, the observations of \cite{wen2016joint} are utilised to construct the priors; $\boldsymbol{\beta}_1=(\beta_{11},\beta_{12})'\sim N_{2}\big(\bigl(\begin{smallmatrix}
0.995\\ 
0.146
\end{smallmatrix}\bigr),\bigl(\begin{smallmatrix}
0.113^{2} & 0\\ 
0 & 0.030^{2} \end{smallmatrix}\bigr)\big)$, $\boldsymbol{\beta}_2=(\beta_{21},\beta_{22})'\sim N_{2}\big(\bigl(\begin{smallmatrix}
1.043\\ 
-0.043
\end{smallmatrix}\bigr),\bigl(\begin{smallmatrix}
0.104^{2} & 0\\ 
0 & 0.030^{2} \end{smallmatrix}\bigr)\big)$ and $\psi^{*}\sim N(0.229,0.122^{2})$.
Furthermore, let $\boldsymbol{\phi}^{*} \sim N_{13}(\boldsymbol{\varphi},\textup{I}_{13})$, where $\boldsymbol{\varphi}=(\varphi_1,\dots,\varphi_{13})$ is estimated using the relation $\hat{\Lambda}_{1}(v_d)=\sum_{d=1}^{13}e^{\varphi_{d}}\Delta_{d}(v_d);d=1,\dots,13$. As the prior of $\boldsymbol{\nu}$, 
$N_{13}(\boldsymbol{\vartheta},\boldsymbol{\Sigma}_{13}(0.2))$ is assigned, with $\vartheta_d=\log\left(-\log\left(\frac{e^{-\hat{\Lambda}_2(v_d)}}{e^{-\hat{\Lambda}_2(v_{d-1})}}\right) \right); d=1,\dots,13$.  

Upon the suitable modification of the data and selection of the appropriate priors, the adaptive MH algorithm with 100,000 iterations is performed retaining every $60^{th}$ sample post burn-in of 20,000 iterations. Each MCMC iteration lasts approximately 0.0104 s. MCMC mixing and convergence diagnostics provided in Appendix \ref{B2} indicate well-converged chains.

\setlength{\tabcolsep}{2pt} % Default value: 6pt
\renewcommand{\arraystretch}{0.9} % Default value: 1
% \usepackage{multirow}
\begin{table}[h]
\centering{
\caption{
Summary of Bayes estimates for the fracture-osteoporosis data }
\label{BCsumm}
% \usepackage{multirow}
% \usepackage{multirow}
\begin{tabular}{|c|cc|cc|c|}
\hline
\multirow{2}{*}{} & \multicolumn{2}{c|}{Mean number of fractures}         & \multicolumn{2}{c|}{Hazard due to osteoporosis}  & \multirow{2}{*}{$\psi$} \\ \cline{2-5}
                  & \multicolumn{1}{c|}{$\beta_{11}$} & $\beta_{12}$ & \multicolumn{1}{c|}{$\beta_{21}$} & $\beta_{22}$ &                         \\ \hline
Estimate          & \multicolumn{1}{c|}{ 0.8371 }           &0.0528             & \multicolumn{1}{c|}{1.0204}             &   -0.0179          &      0.9996                   \\ \hline
PSD               & \multicolumn{1}{c|}{0.0738}             &  0.0214           & \multicolumn{1}{c|}{0.0669}             &   0.0208           &     0.0882                    \\ \hline
BCI               & \multicolumn{1}{c|}{(0.6830,0.9795)}             & (0.0118,0.0949)             & \multicolumn{1}{c|}{(0.8881,1.1479)}             &  (-0.0563,0.0206)            &     (0.8349,1.1819)                    \\ \hline
\end{tabular}}
\end{table}

\begin{figure}[h!]
\centering
\includegraphics[width=.45\textwidth]{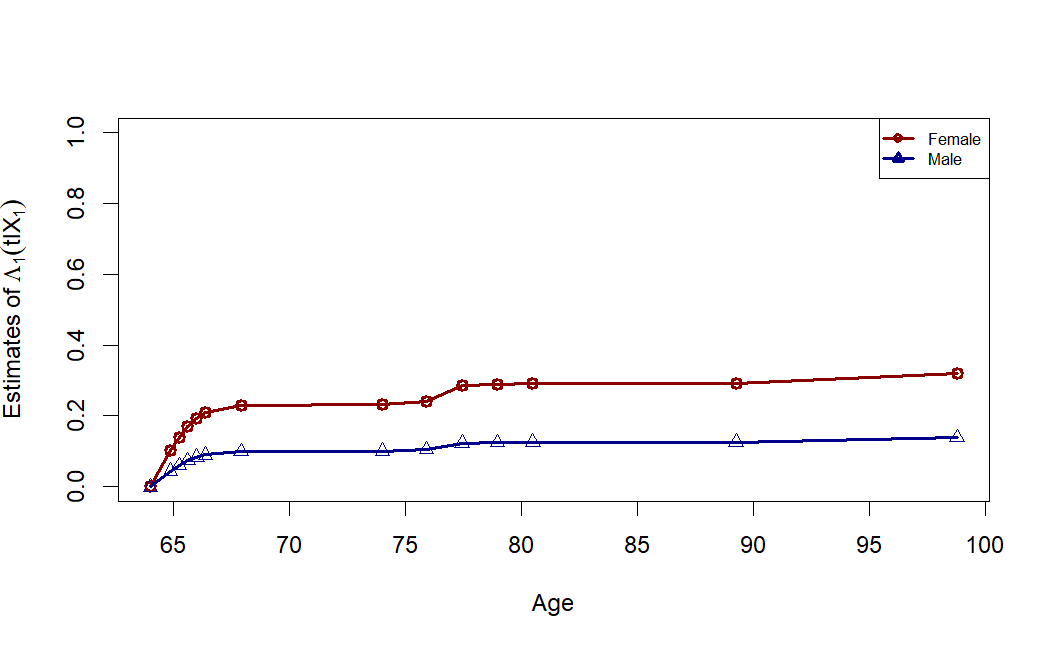}
\includegraphics[width=.45\textwidth]{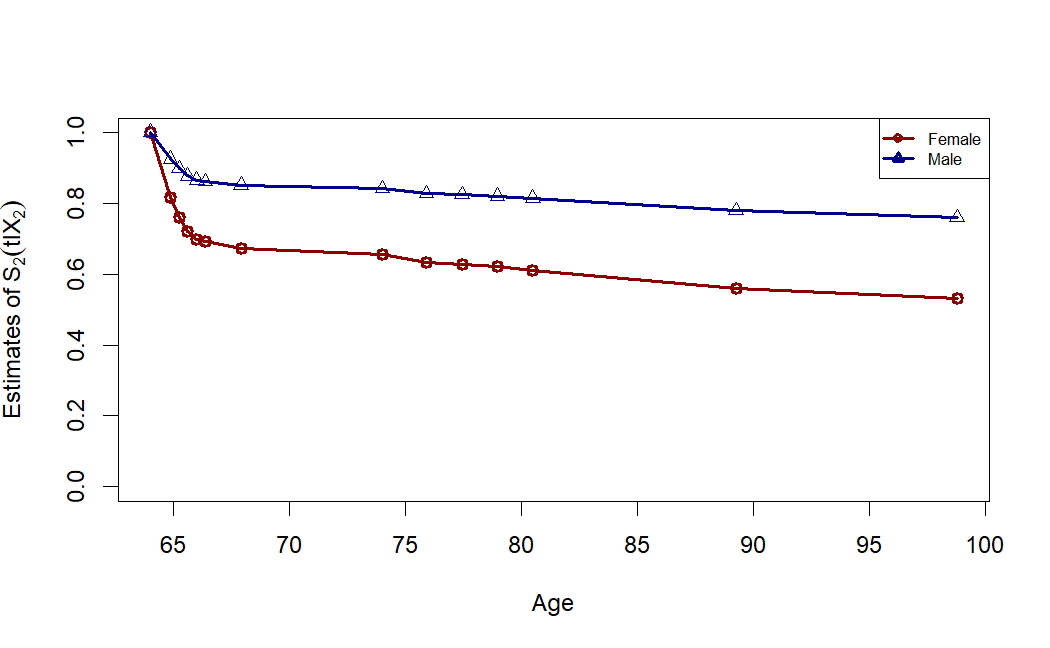}
\caption{Estimates of $\Lambda_1(t|\textbf{X}_1)$ and $S_2(t|\textbf{X}_2)$}
\label{ostmargi}
\end{figure}

It is evident from the posterior summary presented in Table \ref{BCsumm} that gender is a significant risk factor for both the incidence of fractures as well as the development of osteoporosis, since the BCIs of $\beta_{11}$ and $\beta_{21}$ do not contain zero. The non-negative values of these estimates further suggest that females are more vulnerable to the occurrence of both the events. The BCI of $\beta_{12}$ doesn't contain zero, whereas the BCI of $\beta_{22}$ contains zero. These point to the significant impact of tea consumption on fractures, with higher consumption leading to more fractures and statistical insignificance of tea consumption on osteoporosis. The estimate of variance of gamma frailty random variable $\psi$, is also statistically significant, implying a major association between these events. This suggests that people who experience fractures more frequently are more prone to develop osteoporosis. These results are consistent with the observations made by \cite{wen2016joint}. The baseline mean function $\Lambda_{10}(t)$ as well as the baseline cumulative hazard function $\Lambda_{20}(t)$ are estimated over the span of monitoring times, using \eqref{3.6} and \eqref{3.7}. With these estimates, the marginal mean function of fractures' count and the marginal survival function for time to osteoporosis for female and male are estimated keeping tea intake zero, by \eqref{3.8} and \eqref{3.9}, as well as shown in Figure \ref{ostmargi}. The conclusions made are further affirmed by the estimated curves of $\Lambda_1(t|\textbf{X}_1)$, which dominate for females and $S_2(t|\textbf{X}_2)$, which dominate for males.

\begin{figure}[h!]
\centering
\includegraphics[width=6cm,height=4cm]
{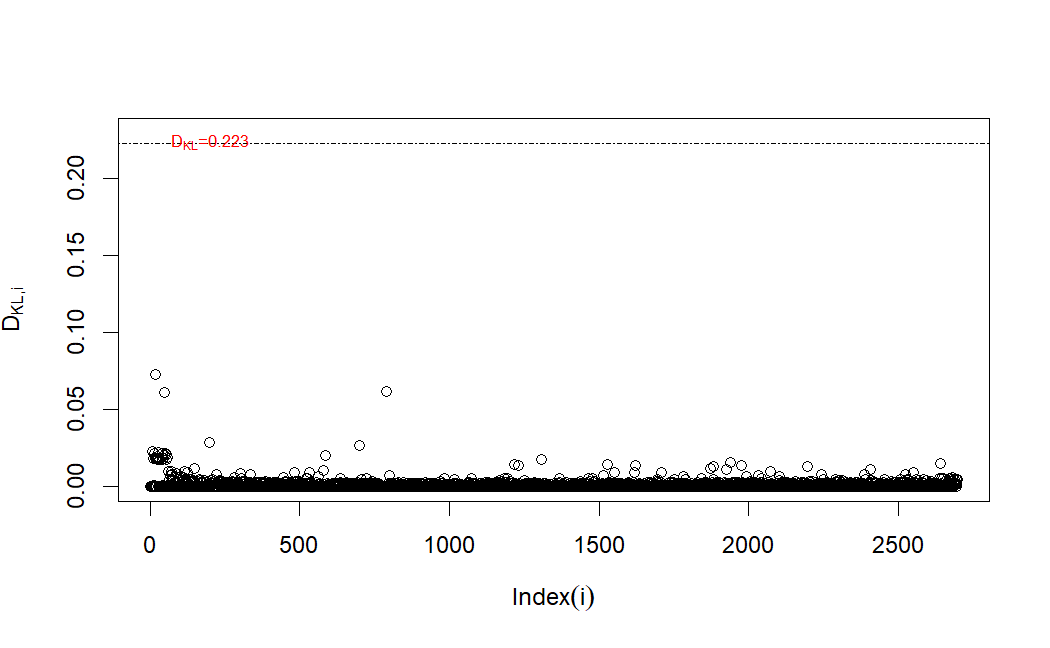}
\caption{
Plot of $D_{\textup{KL},i}$ versus $i$
}
\label{ostkl}
\end{figure}

The model DIC=5475.954 and LPML=-2737.259 are also computed by \eqref{3.15} and \eqref{3.18} respectively.
Model validation is carried out using the KL based Bayesian influence diagnostics derived in \eqref{3.19}. The very small values of $D_{\textup{KL},i}$ plotted against the subject index $i$ for all the subjects under study, as shown in Figure \ref{ostkl}, confirm that the Bayesian model has fitted the data well. Moreover, there are no influential observations, since none of the values are higher than 0.223.

% \begin{figure}[h!]
% \centering
% \includegraphics[width=9cm,height=7cm]{KM_breast.png}
% \caption{
% Kaplan-Meier Survival curve of data set}
% \label{KM_breast}
% \end{figure}

% \begin{figure}[h!]
% \centering
% \includegraphics[width=13cm,height=7cm]{BCfig.png}
% \caption{
% Estimates of survival function}
% \label{PC_DIFFCOV}
% \end{figure}

\section{Conclusion}\label{sec6}
Case I interval censoring happens very commonly while studying epidemics due to their infectious nature.
During several such instances in epidemiological studies, two events; one that recurs results in the occurrence of another non-recurring event. This paper has been an attempt on modelling such data subject to case I interval censoring, namely current count and current status data. A Bayesian approach of estimation is developed for the joint modelling of the data. The approach considers a gamma frailty based semiparametric model, assigns proper priors for the parameters, constructs posterior distribution making use of expected likelihood with regard to frailty, employs an adaptive MH algorithm for posterior computation,
and finally validates the model with a KL based divergence measure. The effectiveness and robustness of the suggested estimation procedure has been verified by comprehensive simulation studies. Illustration with fracture-osteoporosis data suggests the usefulness of the method. 

Current count data are obtained when some recurring event in participants under the study are observed just once and count is recorded. If the recurring events in participants are screened multiple times during a study instead of a single screening and counts are recorded in every such observational window, panel count data are formed, which makes a generalisation of current count data. Under certain instances, some participants may fail to report the actual count in each examination panel. As the result, the only available information on the recurring event of interest is whether it occurred between two successive observation times. This constitutes panel binary data. Bayesian modelling of their mixture, mixture panel count data has not taken place in literature yet. Future work in this regard seems quite promising.

\section*{Acknowledgement(s)}

The first author wishes to acknowledge the financial support of the Council of Scientific \& Industrial Research, Government of India,  via the Junior Research Fellowship scheme under reference No. 09/0239(13499)/2022-EMR-I. The authors would like to extend sincere appreciation to Dr. Chi-Chung Wen, Professor, Department of Mathematics, Tamkang University, Taiwan, for providing the fracture-osteoporosis dataset, used in the analysis.

\section*{Disclosure statement}

There are no conflicts of interest between the authors.

\bibliographystyle{tfs}
\bibliography{ref}

\appendix
\section{Derivations of Bayes Estimators}\label{A}
This appendix details derivations of the Bayes estimators for the parameters; $\phi_{d}$ and $\nu_{d}$ for $d=1,\dots,n'$, $\beta_{1j};j=1,\dots,p$, $\beta_{2j};j=1,\dots,q$, and
 $\psi$.

The marginal posterior density of $\boldsymbol{\phi}^{*}$ is obtained by integrating $\pi ^{*}(\boldsymbol{\Theta}^{*}|\mathbb{D})$;  $\boldsymbol{\Theta}^{*}=(\boldsymbol{\phi^{*}}, \boldsymbol{\nu}, \boldsymbol{\beta_1}, \boldsymbol{\beta_2}, \psi^{*})$ with respect to all parameters except $\boldsymbol{\phi}^{*}$ as 
\begin{equation}\label{A.1}
\begin{aligned}
 \pi_{\boldsymbol{\phi}^{*}}^{*}(\boldsymbol{\phi^{*}}|data)
       &=\idotsint\limits_{\substack{\nu_{d;} \\ d=1,\dots,n'}}\
       \idotsint\limits_{\substack{\beta_{1j;} \\ j=1,\dots,p}}\
\idotsint\limits_{\substack{\beta_{2j;} \\ j=1,\dots,q}}\
\int_{\psi^{*}}
\pi^{*}(\boldsymbol{\Theta}^{*}|\mathbb{D})\\
& \times \prod\limits_{\substack{d=1,\dots,n'}}
d\nu_{d}\prod\limits_{\substack{j=1,\dots,p}}
d\beta_{1j}\prod\limits_{\substack{j=1,\dots,q}}
d\beta_{2j}\,\,d\psi^{*}. 
\end{aligned}
\end{equation}
Therefore the Bayes estimator of $\phi_{d}$ for $d=1,\dots, n'$ is calculated as follows:
\begin{equation}\label{A.2}
 \begin{aligned}
\Tilde{\phi_{d}}
&= E_{\pi_{\phi_{d}^{*}}^{*}}(e^{\phi^{*}_{d}}|\mathbb{D})\\
&= \int_{\phi^{*}_{d}}e^{\phi^{*}_{d}}\pi_{\phi^{*}_{d}}^{*}(\phi^{*}_{d}|\mathbb{D})d\phi^{*}_{d}\\
&=\int_{\phi^{*}_{1}}\int_{\phi^{*}_{2}}\dots \int_{\phi^{*}_{d}}\dots\int_{\phi^{*}_{n'}}e^{\phi^{*}_{d}}\pi_{\boldsymbol{\phi}^{*}}^{*}(\boldsymbol{\phi^{*}}|\mathbb{D})d\phi^{*}_{1}d\phi^{*}_{2}\dots d\phi^{*}_{d}\dots d\phi^{*}_{n'},
\end{aligned}   
\end{equation}
where $\pi_{\phi^{*}_{d}}^{*}(\phi^{*}_{d}|\mathbb{D})$ represents the marginal posterior density of $\phi^{*}_{d};d=1,\dots, n'$.  Analogously, the marginal posterior density of $\boldsymbol{\nu}$ is determined as
\begin{equation}\label{A.3}
\begin{aligned}
 \pi_{\boldsymbol{\nu}}^{*}(\boldsymbol{\nu}|\mathbb{D})
       &=\idotsint\limits_{\substack{\phi^{*}_{d;} \\ d=1,\dots,n'}}\
       \idotsint\limits_{\substack{\beta_{1j;} \\ j=1,\dots,p}}\
\idotsint\limits_{\substack{\beta_{2j;} \\ j=1,\dots,q}}\
\int_{\psi^{*}}
\pi^{*}(\boldsymbol{\Theta}^{*}|\mathbb{D})\\
& \times \prod\limits_{\substack{d=1,\dots,n'}}
d\phi^{*}_{d}\prod\limits_{\substack{j=1,\dots,p}}
d\beta_{1j}\prod\limits_{\substack{j=1,\dots,q}}
d\beta_{2j}\,\,d\psi^{*}. 
\end{aligned}
\end{equation}
Denoting the marginal posterior density of $\nu_{d}$ by $\pi_{\nu_{d}}^{*}(\nu_{d}|\mathbb{D})$, the
Bayes estimator of $\nu_{d}$ for $d=1,\dots,n'$ is
\begin{equation}\label{A.4}
 \begin{aligned}
\Tilde{\nu_{d}}
&= E_{\pi_{\nu_{d}}^{*}}(\nu_{d}|\mathbb{D})\\
 &= \int_{\nu_{d}}\nu_{d}\pi_{\nu_{d}}^{*}(\nu_{d}|\mathbb{D})d\nu_{d}\\
&=\int_{\nu_{1}}\int_{\nu_{2}}\dots \int_{\nu_{d}}\dots\int_{\nu_{n'}}\nu_{d}\pi_{\boldsymbol{\nu}}^{*}(\boldsymbol{\nu}|\mathbb{D})d\nu_{1}\,d\nu_{2}\dots d\nu_{d}\dots d\nu_{n'}.
\end{aligned}   
\end{equation}
$\pi_{\boldsymbol{\beta}_{1}}^{*}(\boldsymbol{\beta_{1}}|\mathbb{D})$, the marginal posterior density of $\boldsymbol{\beta}_{1}$ is 
\begin{equation}\label{A.5}
\begin{aligned}
 \pi_{\beta_{1}}^{*}(\boldsymbol{\beta_{1}}|\mathbb{D})
       &=\idotsint\limits_{\substack{\phi^{*}_{d;} \\ d=1,\dots,n'}}\
       \idotsint\limits_{\substack{\nu_{d;} \\ d=1,\dots,n'}}\
\idotsint\limits_{\substack{\beta_{2j;} \\ j=1,\dots,q}}\
\int_{\psi^{*}}
\pi^{*}(\boldsymbol{\Theta}^{*}|\mathbb{D})\\
& \times \prod\limits_{\substack{d=1,\dots,n'}}
d\phi^{*}_{d}\prod\limits_{\substack{d=1,\dots,n'}}
d\nu_{d}\prod\limits_{\substack{j=1,\dots,q}}
d\beta_{2j}\,\,d\psi^{*}. 
\end{aligned}
\end{equation}
The Bayes estimator of $\beta_{1j}$ for $j=1,\dots,p$ is computed using the marginal posterior density $\pi_{\beta_{1j}}^{*}(\beta_{1j}|\mathbb{D})$ of $\beta_{1j}$ by
\begin{equation}\label{A.6}
 \begin{aligned}
\Tilde{\beta_{1j}}
&= E_{\pi_{\beta_{1j}}^{*}}(\beta_{1j}|\mathbb{D})\\
 &= \int_{\beta_{1j}}\beta_{1j}\pi_{\beta_{1j}}^{*}(\beta_{1j}|\mathbb{D})d\beta_{1j}\\
&=\int_{\beta_{11}}\int_{\beta_{12}}\dots \int_{\beta_{1j}}\dots\int_{\beta_{1p}}\beta_{1j}\pi_{\boldsymbol{\beta}_{1}}^{*}(\boldsymbol{\beta}_{1}|\mathbb{D})d\beta_{11}\,d\beta_{12}\dots d\beta_{1j}\dots d\beta_{1p}.
\end{aligned}   
\end{equation}
$\pi_{\boldsymbol{\beta}_{2}}^{*}(\boldsymbol{\beta_{2}}|\mathbb{D})$, the marginal posterior density of $\boldsymbol{\beta}_{2}$ is 
\begin{equation}\label{A.7}
\begin{aligned}
 \pi_{\beta_{2}}^{*}(\boldsymbol{\beta_{2}}|\mathbb{D})
       &=\idotsint\limits_{\substack{\phi^{*}_{d;} \\ d=1,\dots,n'}}\
       \idotsint\limits_{\substack{\nu_{d;} \\ d=1,\dots,n'}}\
\idotsint\limits_{\substack{\beta_{1j;} \\ j=1,\dots,p}}\
\int_{\psi^{*}}
\pi^{*}(\boldsymbol{\Theta}^{*}|\mathbb{D})\\
& \times \prod\limits_{\substack{d=1,\dots,n'}}
d\phi^{*}_{d}\prod\limits_{\substack{d=1,\dots,n'}}
d\nu_{d}\prod\limits_{\substack{j=1,\dots,p}}
d\beta_{1j}\,\,d\psi^{*}. 
\end{aligned}
\end{equation}
Denoting the marginal posterior density of $\beta_{2j}$ as $\pi_{\beta_{2j}}^{*}(\beta_{2j}|\mathbb{D})$, the
Bayes estimator of $\beta_{2j}$ for $j=1,\dots,q$ is
\begin{equation}\label{A.8}
 \begin{aligned}
\Tilde{\beta_{2j}}
&= E_{\pi_{\beta_{2j}}^{*}}(\beta_{2j}|\mathbb{D})\\
 &= \int_{\beta_{2j}}\beta_{2j}\pi_{\beta_{2j}}^{*}(\beta_{2j}|\mathbb{D})d\beta_{2j}\\
&=\int_{\beta_{21}}\int_{\beta_{22}}\dots \int_{\beta_{2j}}\dots\int_{\beta_{2q}}\beta_{2j}\pi_{\boldsymbol{\beta}_{2}}^{*}(\boldsymbol{\beta}_{2}|\mathbb{D})d\beta_{21}\,d\beta_{22}\dots d\beta_{2j}\dots d\beta_{2q}.
\end{aligned}   
\end{equation}
The Bayes estimator of $\psi$ is
\begin{equation}\label{A.9}
 \begin{aligned}
\Tilde{\psi}
&= E_{\pi_{\psi^{*}}^{*}}(e^{\psi^{*}}|\mathbb{D})\\
 &= \int_{\psi^{*}}e^{\psi^{*}}\pi_{\psi^{*}}^{*}(\psi^{*}|\mathbb{D})d\psi^{*},
\end{aligned}   
\end{equation}
where $\pi_{\psi^{*}}^{*}(\psi^{*}|\mathbb{D})$, the marginal posterior density of $\psi^{*}$ is 
\begin{equation}\label{A.10}
\begin{aligned}
 \pi_{\psi^{*}}^{*}(\psi^{*}|\mathbb{D})
       &=\idotsint\limits_{\substack{\phi^{*}_{d;} \\ d=1,\dots,n'}}\
       \idotsint\limits_{\substack{\nu_{d;} \\ d=1,\dots,n'}}\
\idotsint\limits_{\substack{\beta_{1j;} \\ j=1,\dots,p}}\
\idotsint\limits_{\substack{\beta_{2j;} \\ j=1,\dots,q}}\pi^{*}(\boldsymbol{\Theta}^{*}|\mathbb{D})\\
& \times \prod\limits_{\substack{d=1,\dots,n'}}
d\phi^{*}_{d}\prod\limits_{\substack{d=1,\dots,n'}}
d\nu_{d}\prod\limits_{\substack{j=1,\dots,p}}
d\beta_{1j}\prod\limits_{\substack{j=1,\dots,q}}
d\beta_{2j}. 
\end{aligned}
\end{equation}

\section{MCMC  Convergence and Mixing Diagnostics}\label{B}
The Appendix contains a demonstration of Markov chain convergence for simulation studies and real data analysis. Autocorrelation function (ACF) plots, trace plots, and posterior histograms are analysed as graphical checks. Details are also provided on Gelman-Rubin diagnostics, effective sample sizes (ESS), and acceptance rate.

\subsection{Simulation Studies}\label{B1}
Consider $(\beta_{11},\beta_{21},\psi)=(-1,-1.1,1.2)$ under scenario (1).
With a randomly generated dataset, 100,000 MCMC simulations are done. As burn-in, 10,000 samples are removed and the remaining ones are thinned, keeping only multiples of 30. 

\begin{figure}[h!]
\centering
\includegraphics[width=.3\textwidth]{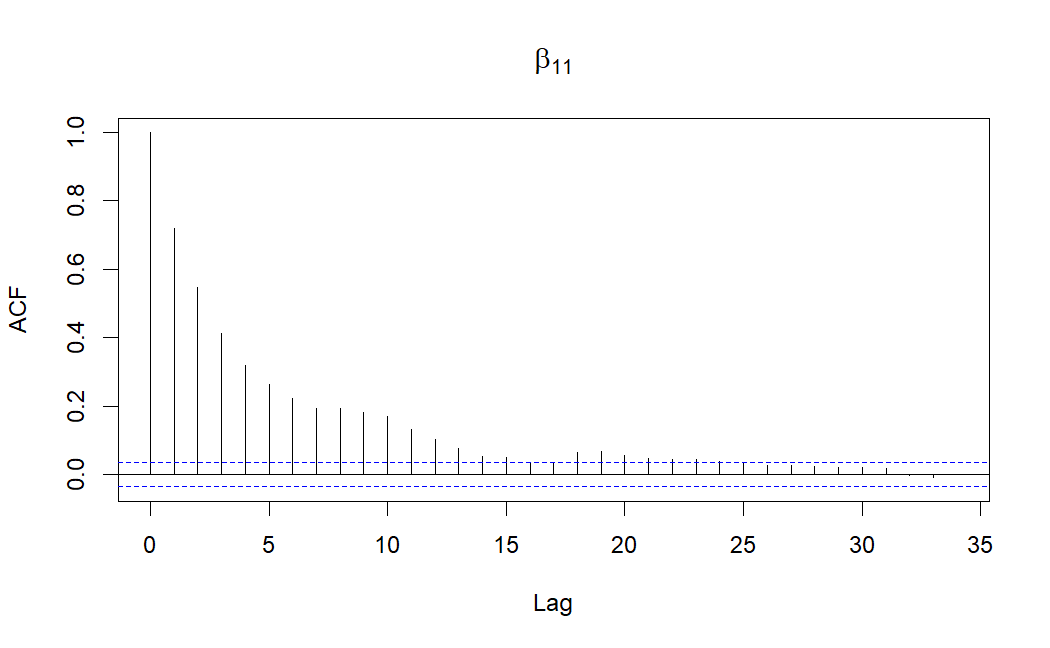}\quad
\includegraphics[width=.3\textwidth]{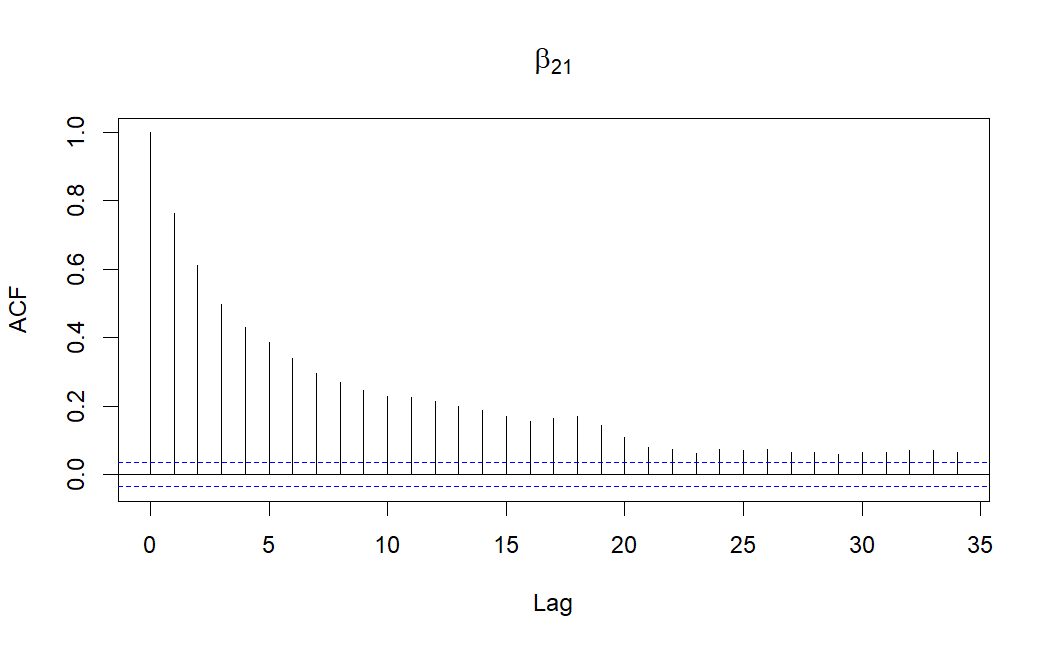}\quad
\includegraphics[width=.3\textwidth]{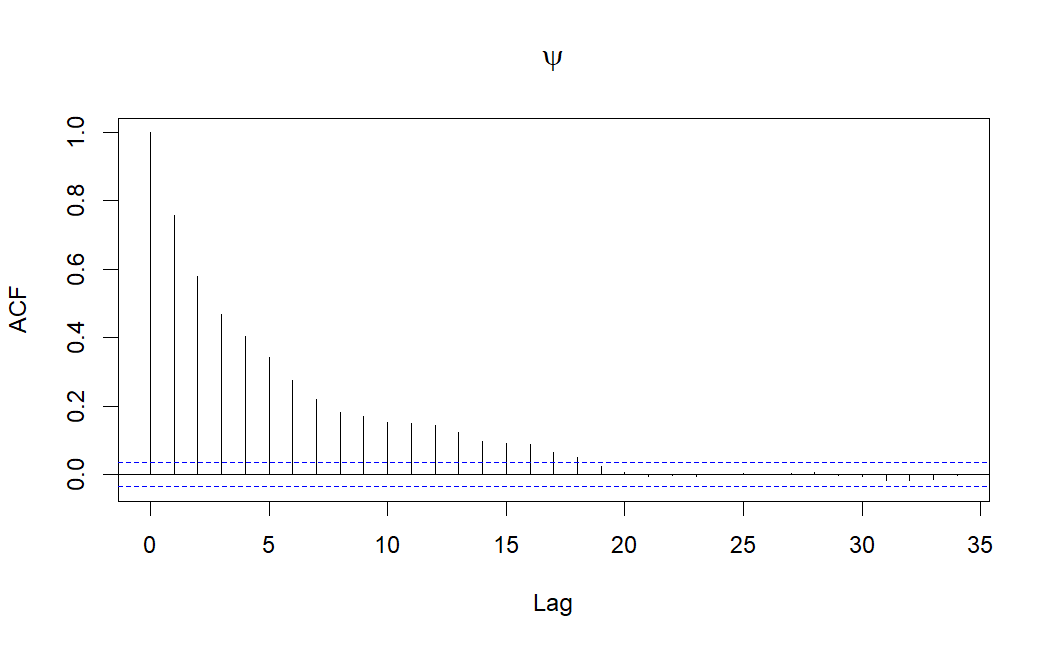}
\caption{ACF plots of parameters when ($\beta_{11},\beta_{21},\psi)=(-1,-1.1,1.2)$}
\label{acf_sim}
\end{figure}

\begin{figure}[h!]
\centering
\includegraphics[width=.3\textwidth]{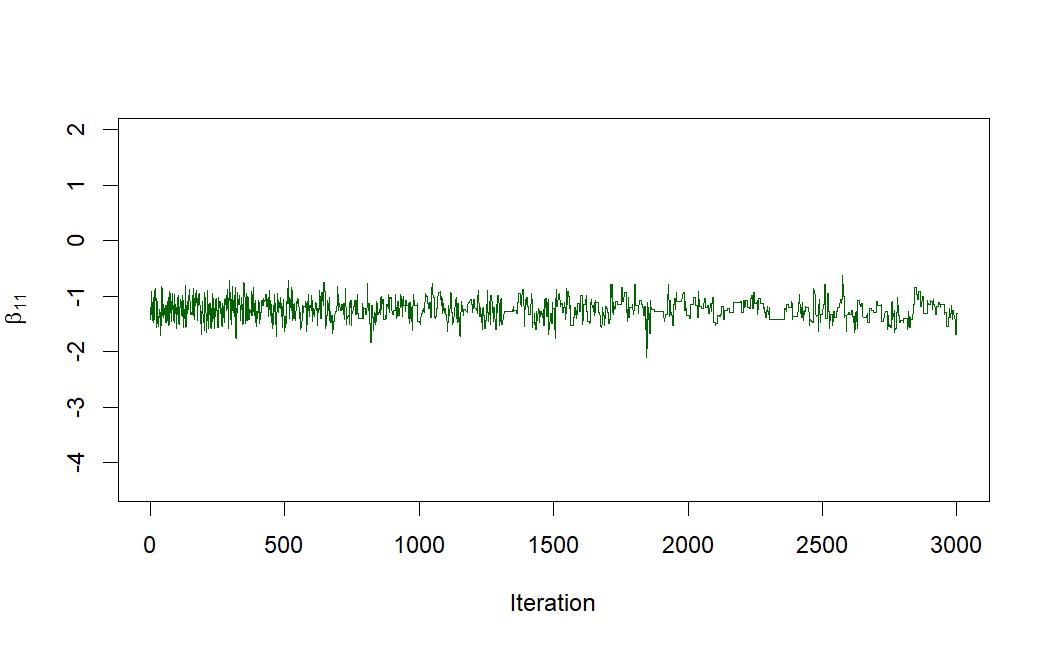}\quad
\includegraphics[width=.3\textwidth]{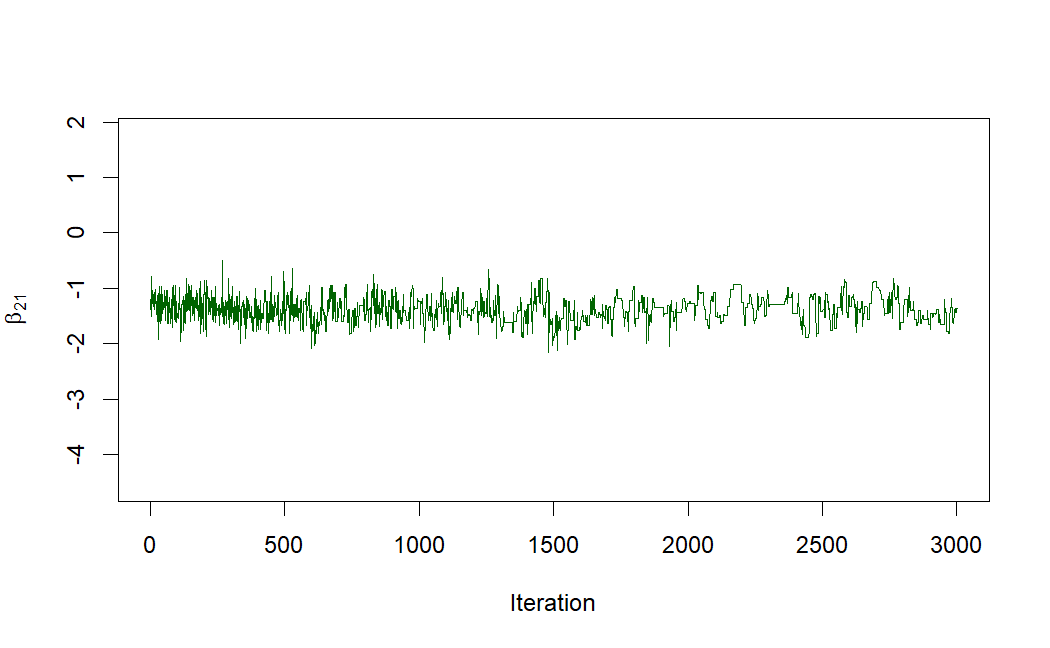}\quad
\includegraphics[width=.3\textwidth]{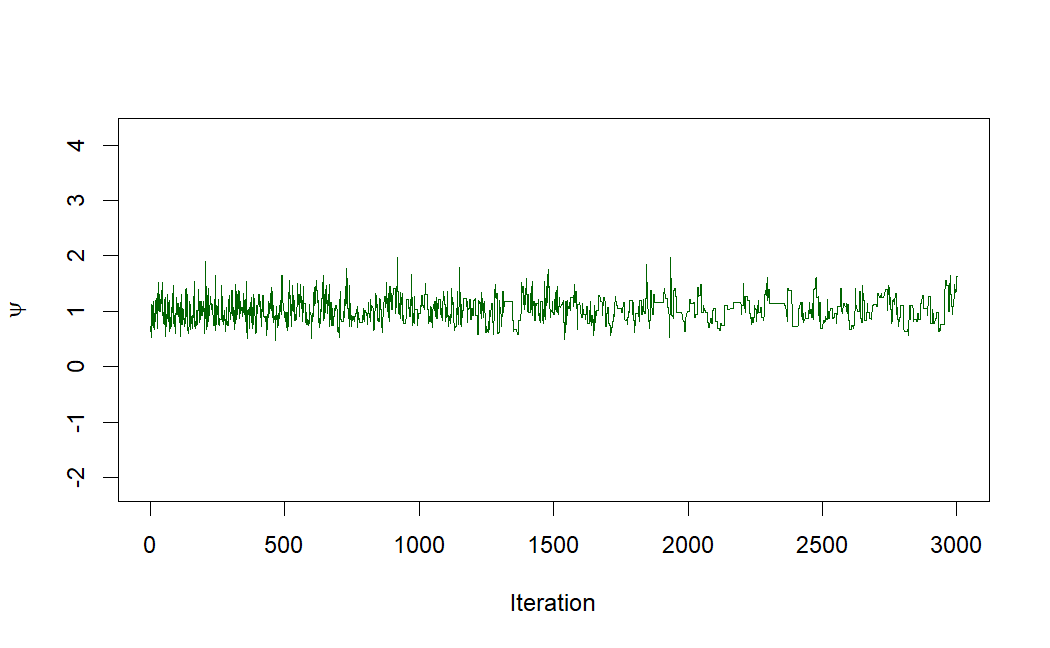}
\caption{Trace plots of parameters when ($\beta_{11},\beta_{21},\psi)=(-1,-1.1,1.2)$}
\label{trace_sim}
\end{figure}

\begin{figure}[h!]
\centering
\includegraphics[width=.3\textwidth]{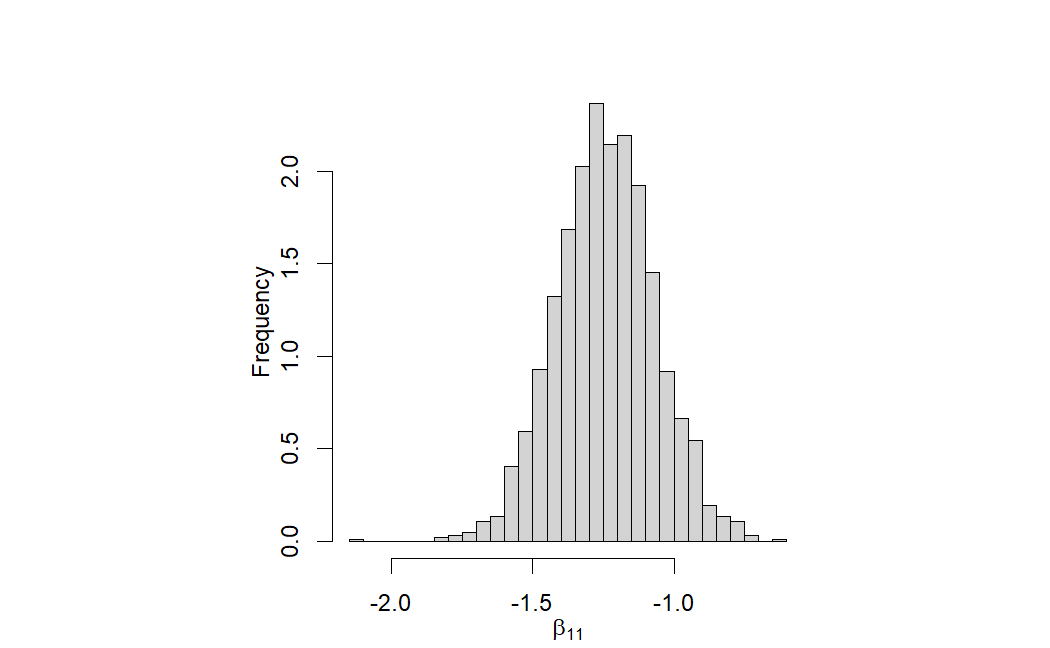}\quad
\includegraphics[width=.3\textwidth]{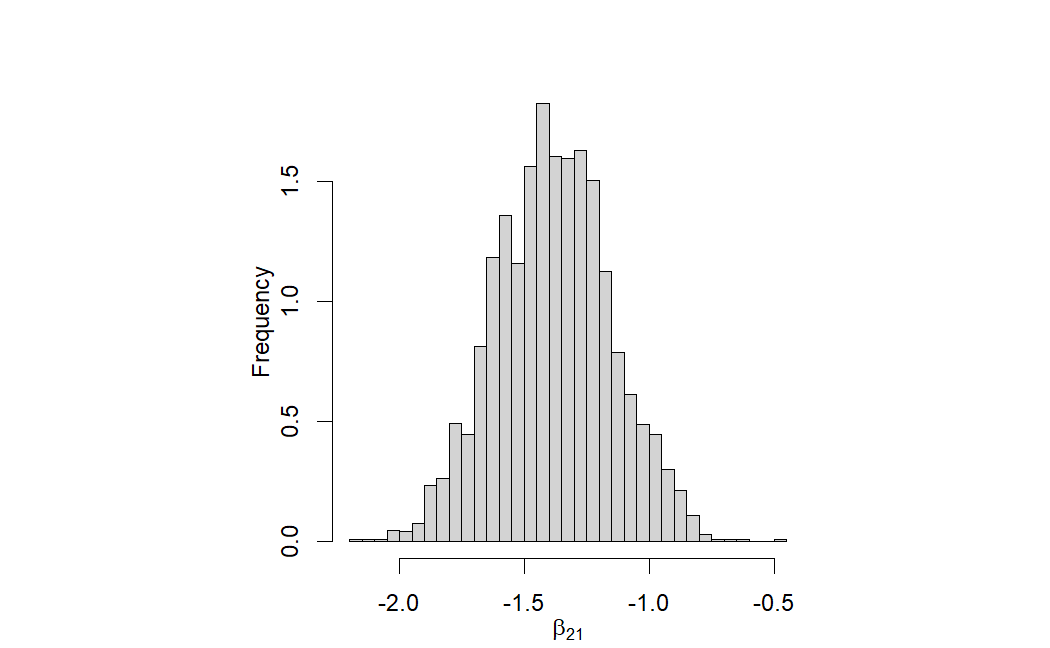}\quad
\includegraphics[width=.3\textwidth]{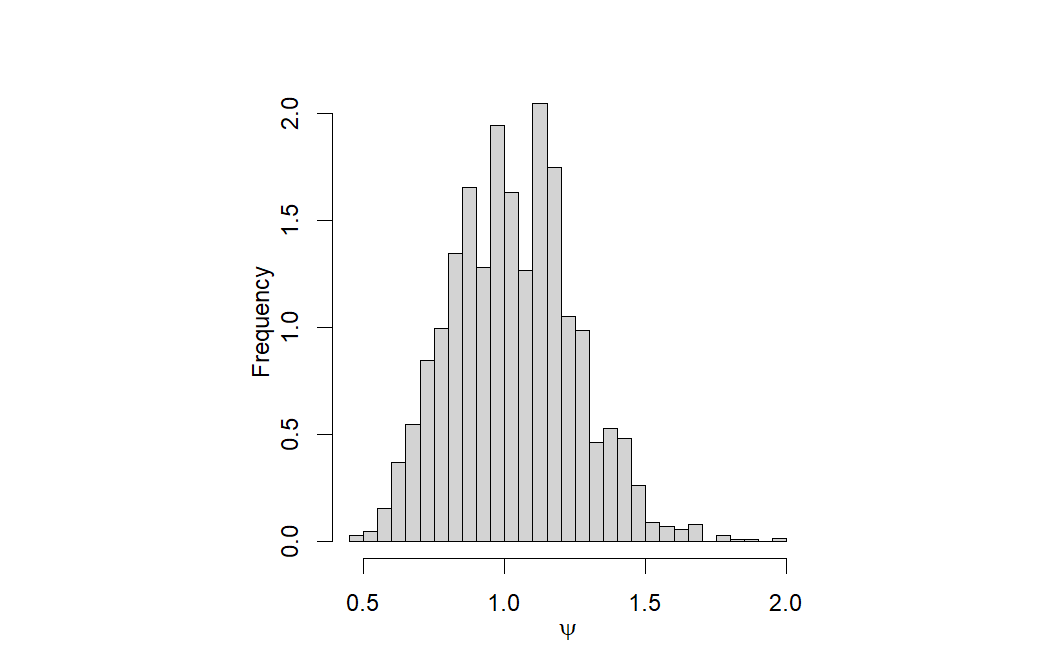}
\caption{Posterior histograms of parameters when ($\beta_{11},\beta_{21},\psi)=(-1,-1.1,1.2)$}
\label{pd_sim}
\end{figure}

ACF plots for $\beta_{11},\beta_{21}$, and $\psi$ in Figure \ref{acf_sim} show chain autocorrelation, where low values indicate good convergence. The plots exhibit rapid autocorrelation decay, signifying well-behaved Markov chains. Trace plots in Figure \ref{trace_sim} reveal MCMC sample progression with random fluctuations around a central value, indicating well-mixed samples. Posterior histograms in Figure \ref{pd_sim} have consistent shapes and narrow peaks, showing convergence and low uncertainty. The Gelman-Rubin diagnostic based on 10 chains, confirms convergence, with potential scale reduction factors close to 1 for $\beta_{11}$, $
\beta_{21}$, and $\psi$. Effective sample sizes for $\beta_{11}$, $\beta_{21}$, and $\psi$ are 261, 182, and 254 respectively. Each MCMC iteration takes about 0.0032 s, with an acceptance rate of 0.0294.

\subsection{Fracture-Osteoporosis Data Analysis }\label{B2}

\begin{figure}[h!]
\centering
\includegraphics[width=.3\textwidth]{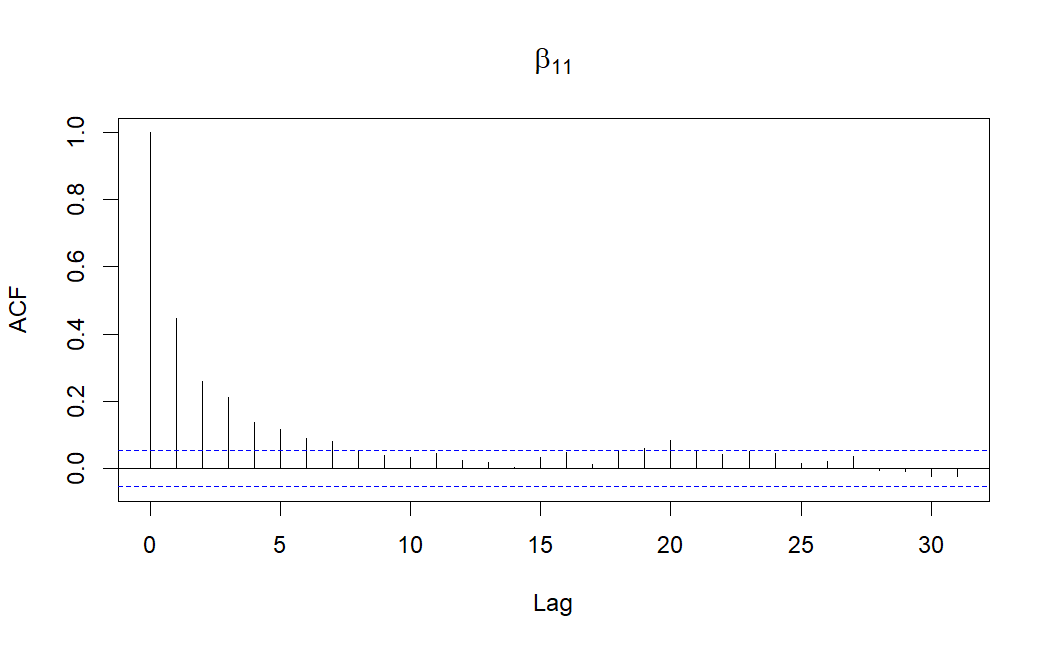}\quad
\includegraphics[width=.3\textwidth]{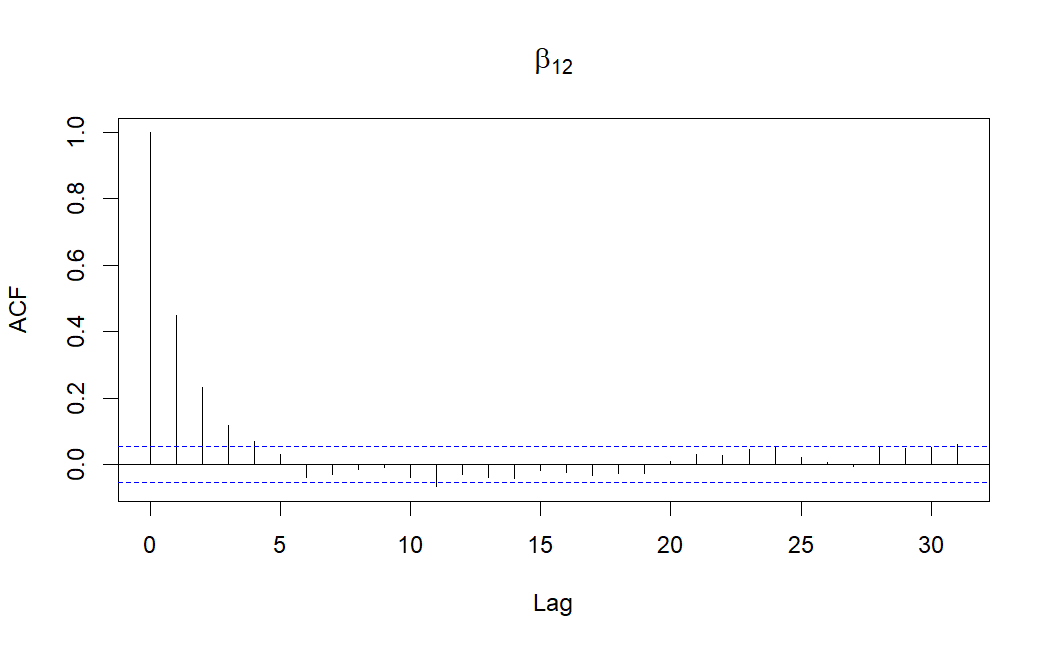}\quad
\includegraphics[width=.3\textwidth]{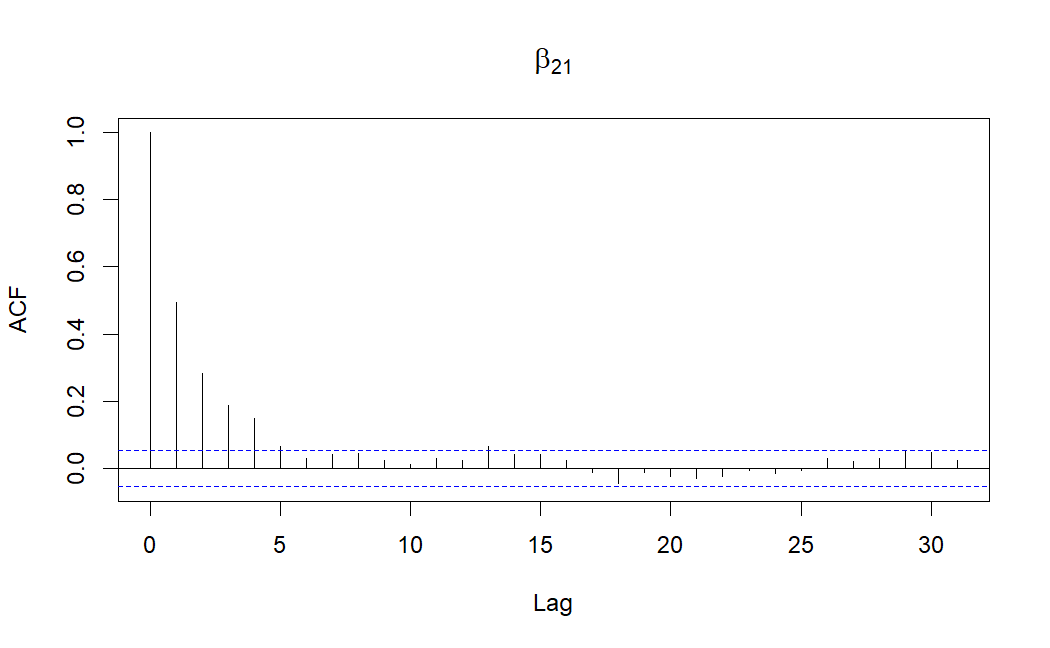}
\medskip
\includegraphics[width=.3\textwidth]{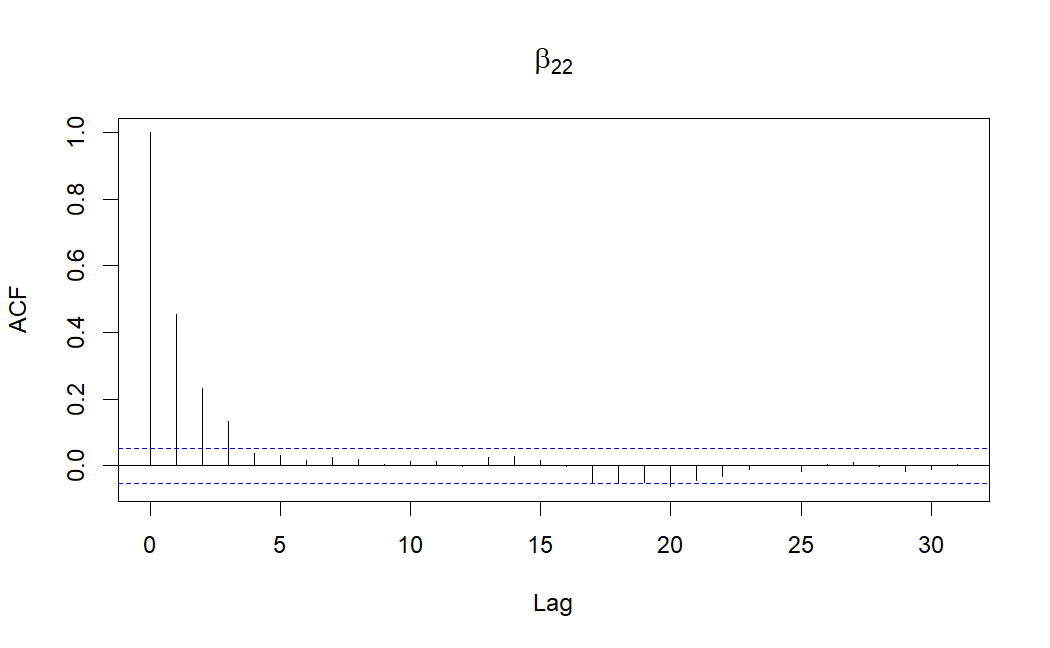}\quad
\includegraphics[width=.3\textwidth]{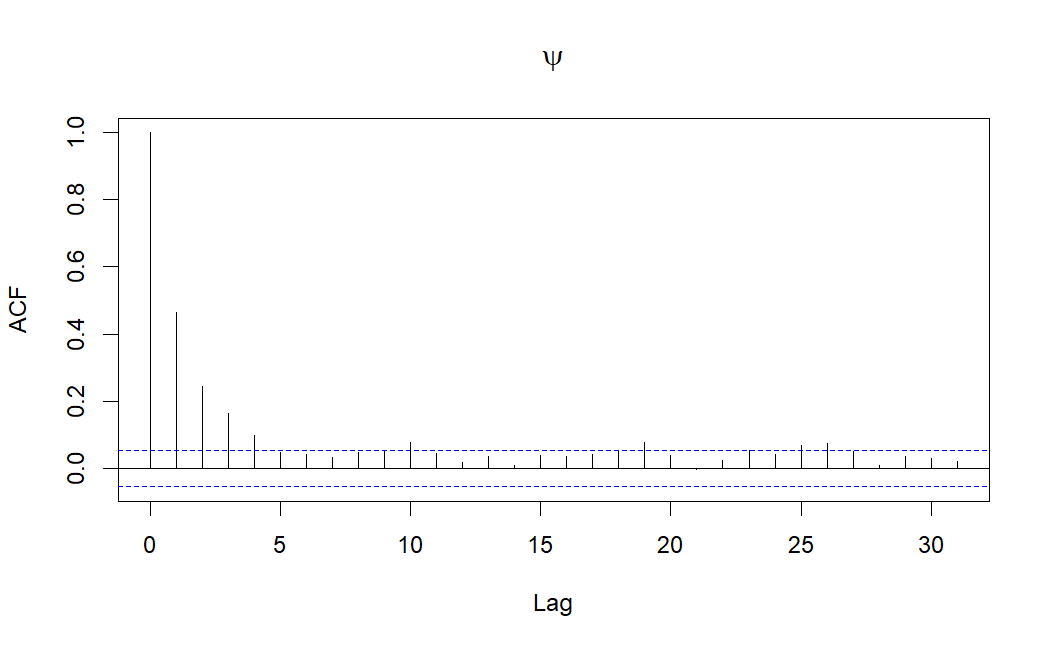}
\caption{ACF plots for $(\beta_{11},\beta_{12},\beta_{21},\beta_{22},\psi)$: Fracture-osteoporosis data analysis}
\label{ost_acf}
\end{figure}

\begin{figure}[h!]
\centering
\includegraphics[width=.3\textwidth]{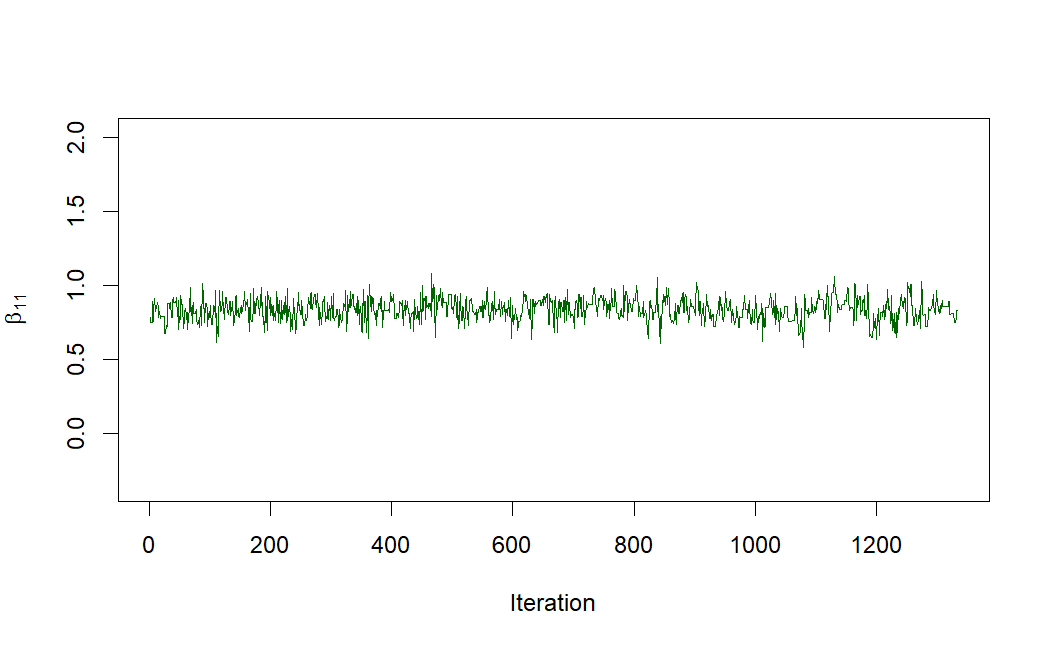}\quad
\includegraphics[width=.3\textwidth]{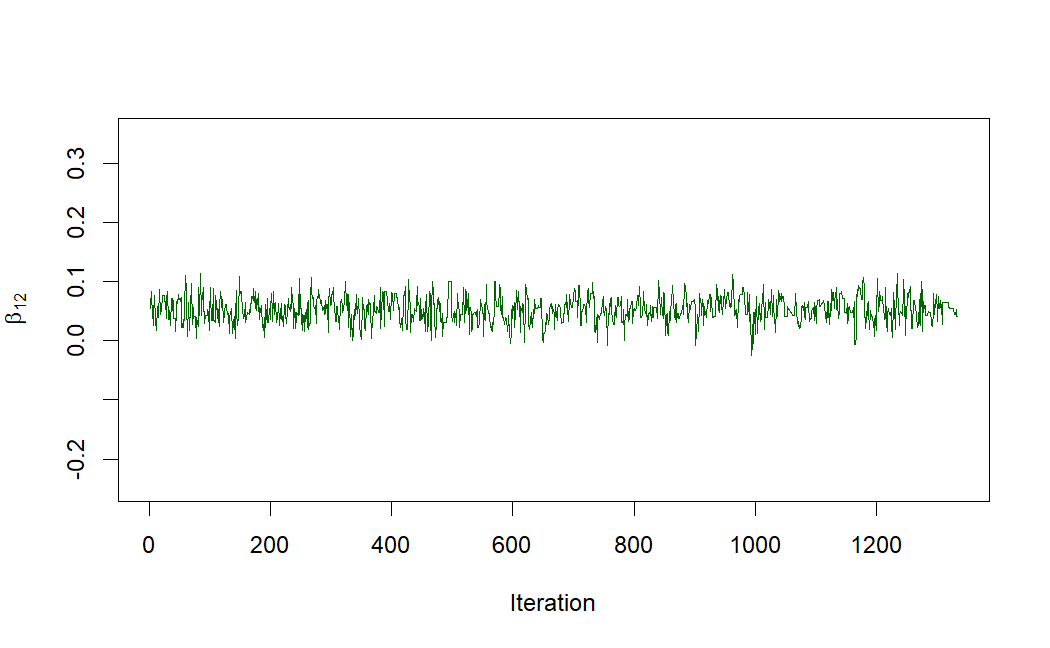}\quad
\includegraphics[width=.3\textwidth]{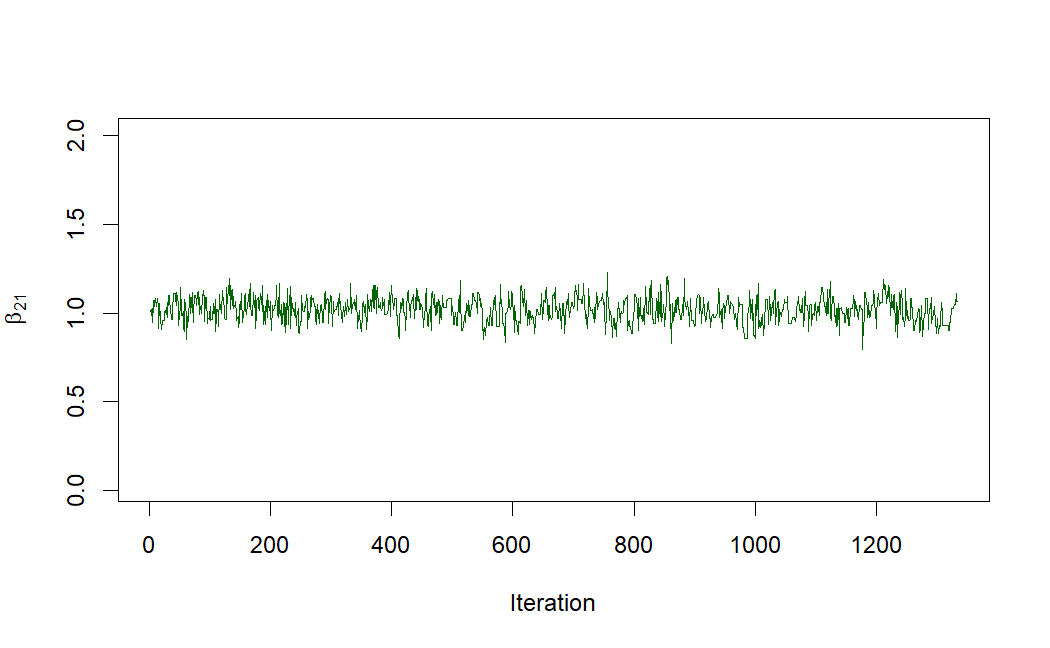}
\medskip
\includegraphics[width=.3\textwidth]{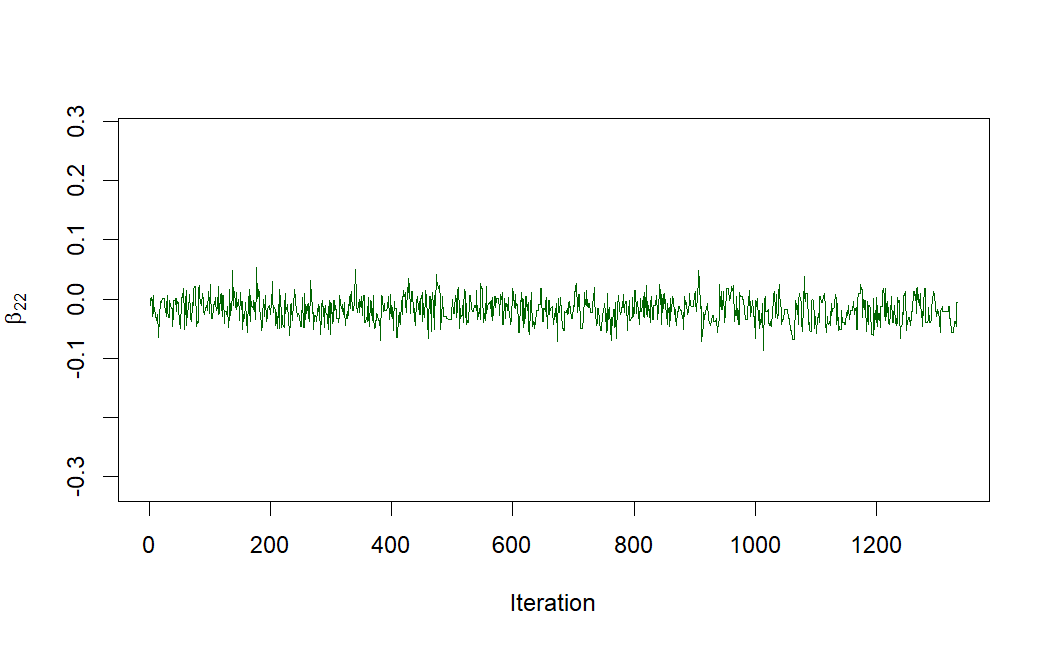}\quad
\includegraphics[width=.3\textwidth]{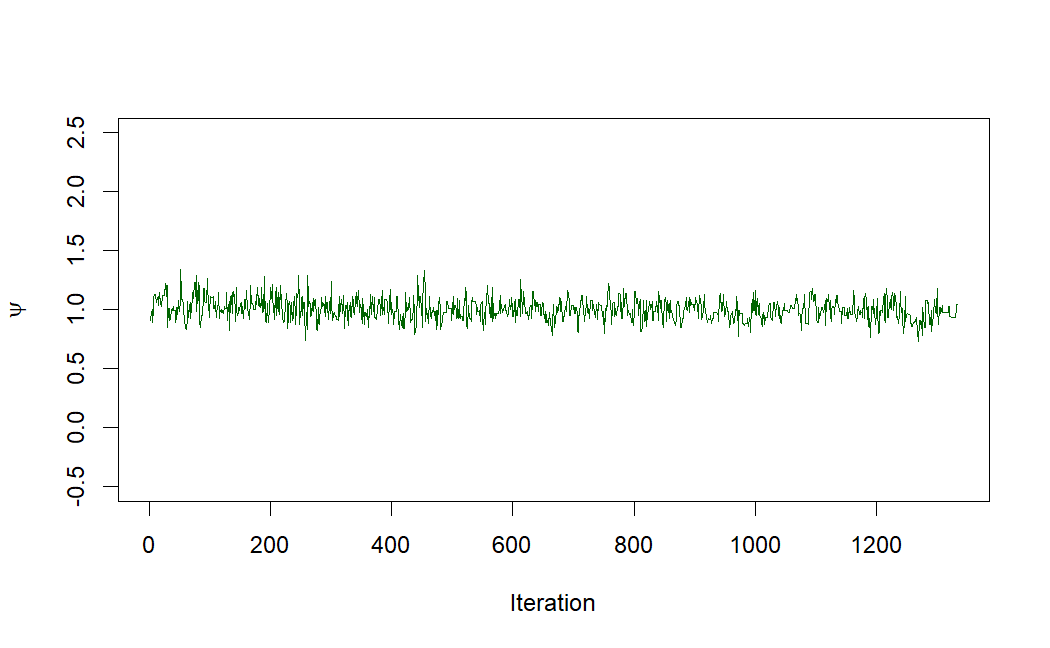}
\caption{Trace plots for $(\beta_{11},\beta_{12},\beta_{21},\beta_{22},\psi)$: Fracture-osteoporosis data analysis}
\label{ost_trace}
\end{figure}

\begin{figure}[h!]
\centering
\includegraphics[width=.3\textwidth]{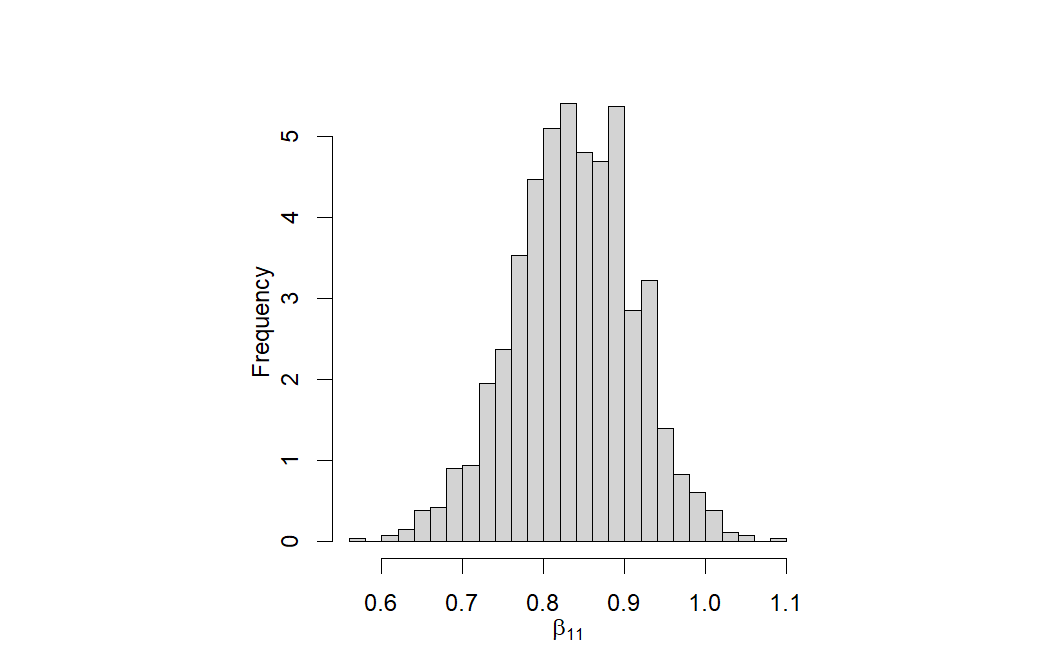}\quad
\includegraphics[width=.3\textwidth]{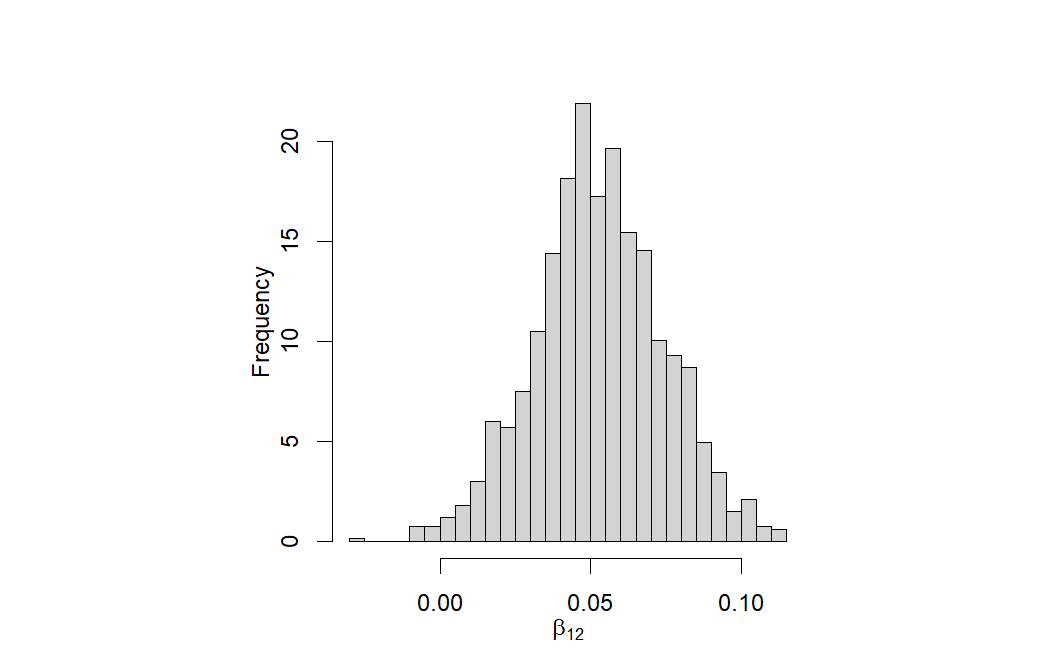}\quad
\includegraphics[width=.3\textwidth]{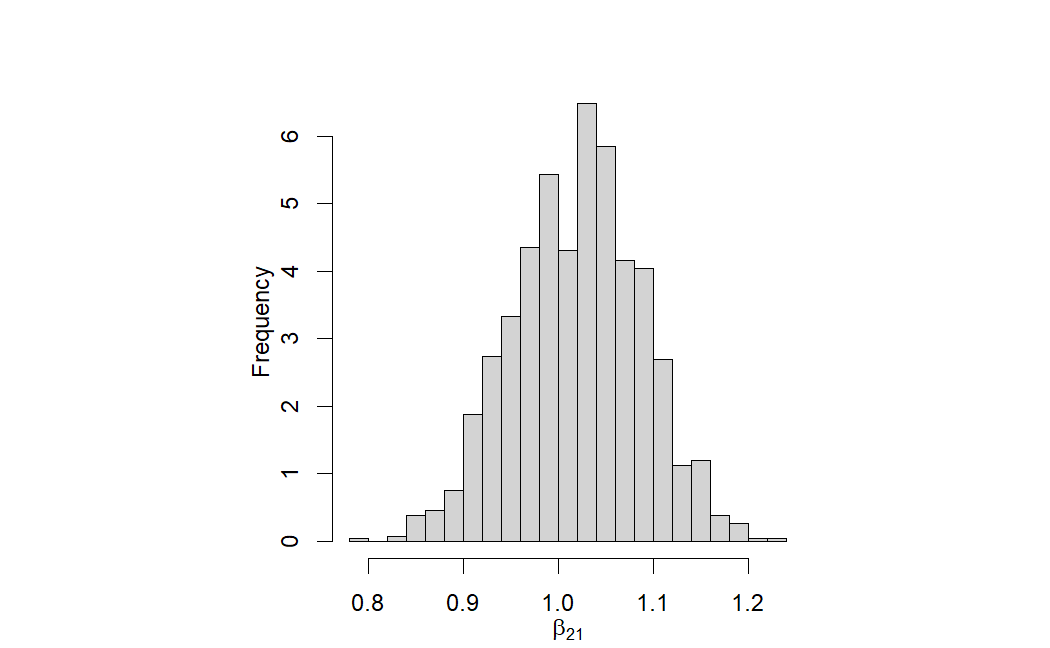}
\medskip
\includegraphics[width=.3\textwidth]{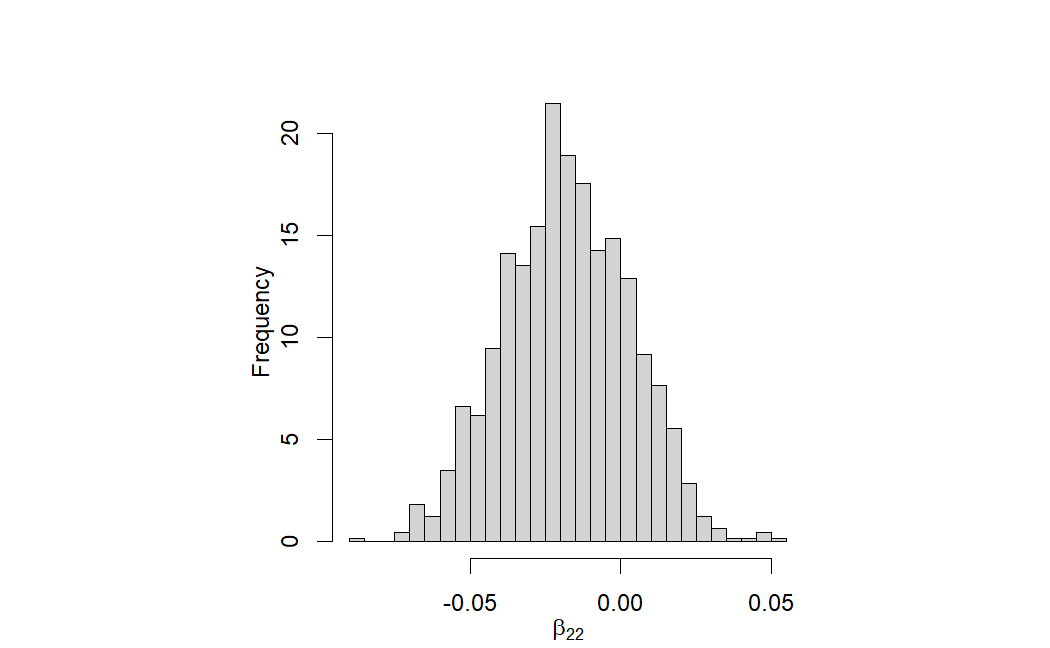}\quad
\includegraphics[width=.3\textwidth]{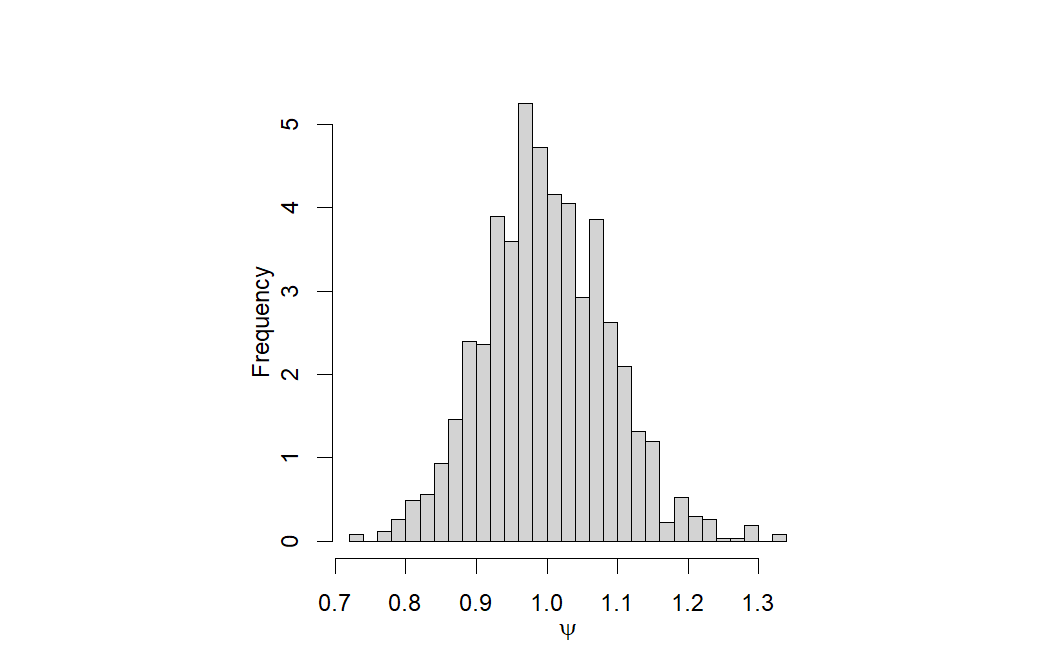}
\caption{Posterior histograms for $(\beta_{11},\beta_{12},\beta_{21},\beta_{22},\psi)$:  Fracture-osteoporosis data analysis}
\label{ost_ph}
\end{figure}

Markov chain diagnostics uses 100,000 MCMC samples for the analysis of fracture-osteoporosis data; 20,000 of these samples are used as burn-in samples and only multiples of 60 are retained. Different graphical diagnostics for $\beta_{11},\beta_{12},\beta_{21},\beta_{22}$, and $\psi$ shown in Figures \ref{ost_acf}, \ref{ost_trace}, and \ref{ost_ph} indicate satisfactory convergence of the chains. Further evidence comes from Gelman-Rubin diagnostics values near to 1. ESS for $\beta_{11}$, $\beta_{12}$, $\beta_{21}$, $\beta_{22}$,  and  $\psi$ are 184, 239, 203, 260, and 185 respectively. There is an acceptance rate of 0.0364 and computing time for each iteration is 0.0104 s.

\end{document}